# Active and Passive Hybrid Detection Method of Power CPS False Data Injection Attacks Based on Improved AKF and GRU-CNN

**ZHAOYANG QU[1,3], XIAOYONG BO[1,2,3], TONG YU[4], YAOWEI LIU[5], YUNCHANG DONG[1,3], ZHONGFENG KAN[6], LEI WANG[1,3], YANG LI[1]**

[1]School of Electrical Engineering, Northeast Electric Power University, Jilin 132012, China
[2]Electrical and Information Engineering College, Jilin Agricultural Science and Technology University, Jilin 132101, China
[3]Jilin Province Engineering Technology Research Center of Power Big Data Intelligent Processing, Jilin 132012, China
[4]Guangxi Power Grid Co., Ltd. Electric Power Research Institute, Nanning 530013, China
[5]State Grid Jilin Electric Power Co., Ltd., Changchun 130012, China
[6]State Grid Jilin Electric Power Co., Ltd. Jilin Power Supply Company, Jilin 132011, China

Corresponding author: Xiaoyong Bo (110647710@qq.com).

This work was supported in part by the Key Projects of the National Natural Science Foundation of China under Grant 51437003, and in part by the Jilin Science and Technology Development Plan Project of China under Grant 20180201092GX and 20200401097GX.

**ABSTRACT** With the deep penetration of the new generation of information technology in the power system, the power system has gradually evolved into a highly coupled cyber-physical systems (CPS), and false data injection attacks (FDIAs) is the most threatening attack among many power CPS network attacks. Aiming at the problem that the existing knowledge-driven detection process of FDIAs has been in a passive detection state for a long time and ignores the advantages of data-driven active capture of features, an active and passive hybrid detection method for power CPS FDIAs with improved adaptive Kalman filter (AKF) and convolutional neural networks (CNN) is proposed in this paper. First, it analyzes the shortcomings of the traditional AKF algorithm in terms of filtering divergence and calculation speed. The state estimation algorithm based on non-negative positive definite adaptive Kalman filter (NDAKF) is improved, and FDIAs passive detection method with similarity Euclidean distance detection and residual detection as the core is constructed. Then, combined with the advantages of gate recurrent unit (GRU) and CNN in terms of temporal memory and feature expression ability, an active detection method of FDIAs based on GRU-CNN hybrid neural network is designed. Finally, the results of joint knowledge-driven and data-driven parallel detection define a mixed fixed calculation formula, and the active and passive hybrid detection method of FDIAs is established considering the characteristic constraints of the parallel mode. The simulation system example of power CPS FDIAs verifies the effectiveness and accuracy of the method proposed in this paper.

## I. INTRODUCTION

With the deep penetration of the new generation of information technology in the power system [1-4], a large number of electrical equipments, data acquisition devices, and terminal computing units are interconnected through two physical networks of the power grid and the communication network. The traditional power system with physical equipment as the core has gradually evolved into a highly coupled cyber-physical systems (CPS) [5-8]. As a monitoring and control system to ensure the safe and stable operation of the power grid, the power CPS realizes the supervisory control and data acquisition (SCADA) system and the phasor measurement unit (PMU) measurement data and control instruction transmission, and energy management system (EMS) data processing, analysis and decision-making, it can be seen that the power system is more and more dependent on CPS [9-12]. But at the same time, with the close integration of power CPS computing systems, communication networks, sensor networks, control systems, and physical systems, more frequent information interaction has also increased the risk of network attack on the power grid. In 2010, the Iranian nuclear facility was attacked by the Stuxnet virus, the Ukrainian power grid was attacked by the BlackEnergy virus in 2015, the U.S. wind farm was attacked by the ARP cache virus in 2017, and the Venezuelan hydropower plant control center was attacked by a network in 2019 [13-14]. They are typical cases of large-scale regional blackout caused by damage to the power CPS.

False Data Injection Attacks (FDIAs), as one of the more threatening attack methods in many power CPS network attacks, has strong accessibility, interference and concealment. In 2009, a false data injection attacks for power grid state estimation was first proposed by Liu Yao et al [15]. They pointed out that the attacker invads the system from the power CPS information communication network, and obtains the power CPS network parameters and topology, by manipulating the measuring device, constructs false measurement data that satisfies the constraints of state estimation, avoid the bad data detection process, and launch attacks without being noticed by the control center, so that the control center loses its ability to perceive the current system operating state or topology, thereby inducing it produce wrong estimates and issues wrong instructions, which disrupts the normal operation of the power grid [16-18]. Power CPS FDIAs completely invalidate the traditional bad data detection mechanism, posing a serious threat to the robustness of the safe operation of the power grid. Therefore, based on the idea of power CPS fusion, the cyber-physical comprehensive analysis of the FDIAs process is carried out, and the feature detection methods are studied from the knowledge-driven and data-driven perspectives, and the active and passive hybrid detection mechanism of power CPS FDIAs is established. The safe and stable operation of CPS is of great significance.

At present, domestic and foreign research on detection methods for FDIAs of power CPS mainly focuses on state estimation detection methods [19-23], trajectory prediction detection methods [24-26], and artificial intelligence detection methods [16, 27-28, 36]:

In terms of state estimation detection methods, literature [19] proposed a new FDIAs detection and identification method, which uses equivalent measurement transformation to replace the traditional weighted least squares method in the state estimation process, and uses residual detection to identify FDIAs. Literature [20] based on the analysis of the FDIAs process, by protecting a set of sensor measurement values selected by the strategy, independently verifying or measuring the value of the state variable selected by the strategy, and then exploring a FDIAs detection method. Literature [21] considering the robustness of different state estimators, by running multiple robust least squares estimators with different breakdown points in parallel, the network attack detection method for grid state estimation is improved, thereby improving the power CPS state estimation. Literature [22-23] proposed a tolerable FDIAs detection method based on extended distributed state estimation, using graph segmentation method to divide the power grid into multiple subsystems, each subsystem is extended outwards to generate extended subsystems, using the chi-square test to detect the wrong data in each extended subsystem. The false data is obviously different from the normal observation error, thereby improving the sensitivity of detection. The existing state estimation detection methods are passive detection behaviors, and they all use mature algorithms. Although the detection speed is fast and can better reflect the characteristics of power CPS, the detection threshold setting has a greater impact on the detection accuracy, and it is easy to missing and false detection.

In terms of trajectory prediction detection methods, literature [24] extended the approximate DC model to a general linear model, derived a general FDIAs model, and on this basis, the consistency between the predicted measurement value and the received measurement value is tested based on the short-term state prediction method considering the time correlation and the statistical consistency test method. Literature [25] proposed a sequence detector based on the generalized likelihood ratio to solve the FDIAs detection problem. The proposed detector is robust against various attack strategies and the load situation in the power system, and its computational complexity is linearly proportional to the number of measuring devices, which ensures the high-performance characteristics of the detector. Literature [26] proposed a model prediction method based on multi-sensor track fusion, in order to extract the initial relevant information of the attacked oscillation parameters, a Kalman-like particle filter-based smoother is used on each monitoring node, and the smoother is diagonalized into a subsystem to deal with the grid caused by FDIAs Continuous load fluctuations and disturbances. The existing trajectory prediction detection methods belong to passive detection behavior. According to the operation law of the system state and the historical database, the distribution law of the state variable is predicted, and the operation trajectory is matched, which can effectively detect various types of false data, but the computational complexity is high, the detection speed is slow, and it is not suitable for complex systems.

In terms of artificial intelligence detection methods, literature [16] proposed a power grid FDIAs detection method based on XGBoost combined with unscented Kalman filtering, the state quantity obtained from the XGBoost load prediction result and the state quantity obtained from the UKF dynamic state estimation are used for adaptive hybrid prediction, through the central limit theorem to compare the distribution of random variables for FDIAs detection. Literature [27] used deep learning technology to extract FDIAs behavioral characteristics of historical measurement data, and uses the captured features to detect FDIAs in real time. The proposed detection mechanism effectively relaxes the hypothesis of potential attack scenarios has been obtained, and a high detection accuracy has been obtained. Literature [28] used batch processing and online learning algorithms (supervised and semi-supervised) to combine decision-making and feature-level fusion to establish an attack detection model, and analyzed the relationship between the attack scenario and the statistical and geometric attributes of the attack vector used in the learning algorithm to use statistical learning methods to detect unobservable attacks. Literature [36] constructed a FDIAs detection model based on an improved convolutional neural network, and implemented an efficient real-time FDIAs detector based on the proposed model design. The existing artificial intelligence detection methods are active detection behaviors, and their significant advantages are strong computing power and clear framework. However, due to the complexity of the operation mechanism of power CPS, the interpretability of such methods is usually poor.

In general, the current research on the detection methods of power CPS FDIAs is full of flowers, each with its own strengths. Some are from the perspective of state estimation, focusing on detection speed, and some are from the perspective of trajectory prediction, focusing on detection accuracy, others are cutting in from the perspective of artificial intelligence, focusing on computing power and frameworks, but no matter from which point of view, they have their own advantages and disadvantages. For the time being, this paper summarizes the two perspectives of state estimation and trajectory prediction as knowledge-driven method, and the perspective of artificial intelligence as data-driven method, the two methods are comprehensively analyzed. On the one hand, the knowledge-driven method has theoretical support and the results of the analysis are reasonable. Although credible and purely data-driven artificial intelligence methods can dig out data rules, they are difficult to explain from the mechanism to make people convincing, and it takes time to collect new information to complete the update when new scenarios such as system structure changes appear. On the other hand, the interactive process of power CPS is complex and the state space is huge. It is difficult to ensure the comprehensiveness and accuracy of extracting features by relying solely on knowledge-driven analysis. However, data-driven method can capture features that are reflected in data but may be overlooked, mis-simplified, or the mechanism is not yet understood in theoretical analysis. Therefore, it is very necessary to combine knowledge-driven and data-driven detection method. Complementing each other can help improve the efficiency and accuracy of FDIAs detection.

Through the above analysis, based on the improved AKF algorithm and GRU-CNN neural network, an active and passive hybrid detection method for power CPS FDIAs is proposed in this paper. The main contributions of this paper are as follows:

1) Improved the AKF algorithm from a knowledge-driven perspective. First, re-improve the AKF algorithm based on the non-negativity of the positive definite diagonal matrix, predictive measurement state estimation value through NDAKF algorithm, and then use the similarity Euclidean distance detection method and the residual detection method to determine the threshold to realize the passive detection of FDIAs.

2) Designed a GRU-CNN hybrid neural network model from a data-driven perspective. First, analyze the respective advantages of GRU and CNN, and then combine the strong memory of GRU and the strong feature expression ability of CNN to design a GRU-CNN hybrid neural network model, and realize the active detection of FDIAs through model training and algorithm design.

3) The active and passive hybrid detection method is designed from the perspective of parallel mode. The results of knowledge-driven and data-driven parallel detection are comprehensively processed, and the characteristic constraints of parallel modes in the typical joint mode are considered, and the result of the combined operation of the two is taken as the final output result to realize the active and passive hybrid detection of FDIAs.

The remainder of this paper is organized as follows. The core detection architecture and related models are built in Section II. A passive detection method of FDIAs based on improved AKF algorithm is proposed in Section III. An active detection method of FDIAs based on GRU-CNN hybrid neural network is designed in Section IV. The active and passive hybrid detection method of FDIAs is established in Section V. The effectiveness and accuracy of the proposed method are verified and analyzed in Section VI. Finally, a summary of this paper is provided in Section VII.

## II. ACTIVE AND PASSIVE HYBRID DETECTION ARCHITECTURE AND RELATED MODELS FOR POWER CPS FDIAs

### *A. PROBLEM DESCRIPTION*

FDIAs acts on the power CPS, Compared with other attacks that directly damage the operation of the power CPS, and its outstanding feature lies in the concealment of the attack. Based on the acquired system functions, parameter configuration and topology information, the attacker uses the vulnerabilities of the device and the imperfect confidentiality mechanism to gain access to the communication system, and achieves the purpose of tampering with the data by injecting false data.

In the power CPS, the information network adopts the principle of "secure partition, dedicated network, horizontal isolation, and vertical authentication", using high-security, high-reliability, and high-bandwidth optical fiber private network communication, the information network related equipment and channels have a very low probability of being attacked by information. Furthermore, the security protection level of the secondary equipment of the public communication network and the power grid terminal is low, which has become a common network attack intrusion point. Attackers can successfully attack power CPS equipment by attacking these potential intrusion locations, and then using the loopholes and weak security mechanisms that exist in the system. Therefore, the network attack method is mainly aimed at the communication transmission link and secondary terminal equipment, that is, attacking the uplink/downlink communication channel and the remote terminal unit RTU (Remote Terminal Unit) or the phasor measurement unit PMU.

For the uplink communication process, FDIAs can tamper with the measured system parameters or operating information, bypassing the bad data detection mechanism, and causing the data information received by the control center to be inconsistent with the actual system operating state, which affects the control center's decision-making. The uplink communication process is affected by FDIAs, which is the focus of this paper. For the downlink communication process, control commands (such as circuit breaker opening and closing, generator output, etc.) can be tampered with, causing the actuator to execute fake control commands, which has a direct impact on the operation of the power grid.

As shown in Figure 1, it is a typical power CPS FDIAs process. The attacker penetrates the monitoring system through the inherent attack steps designed in advance, and directional tampering with the measurement data to destroy the information integrity of the power CPS, thereby affecting the analysis and decision-making of the upper-level control center. Then issuing control instructions, the system switch and the knife switch will refuse to move or malfunction, which causes serious consequences such as large-scale power outages in the power system.

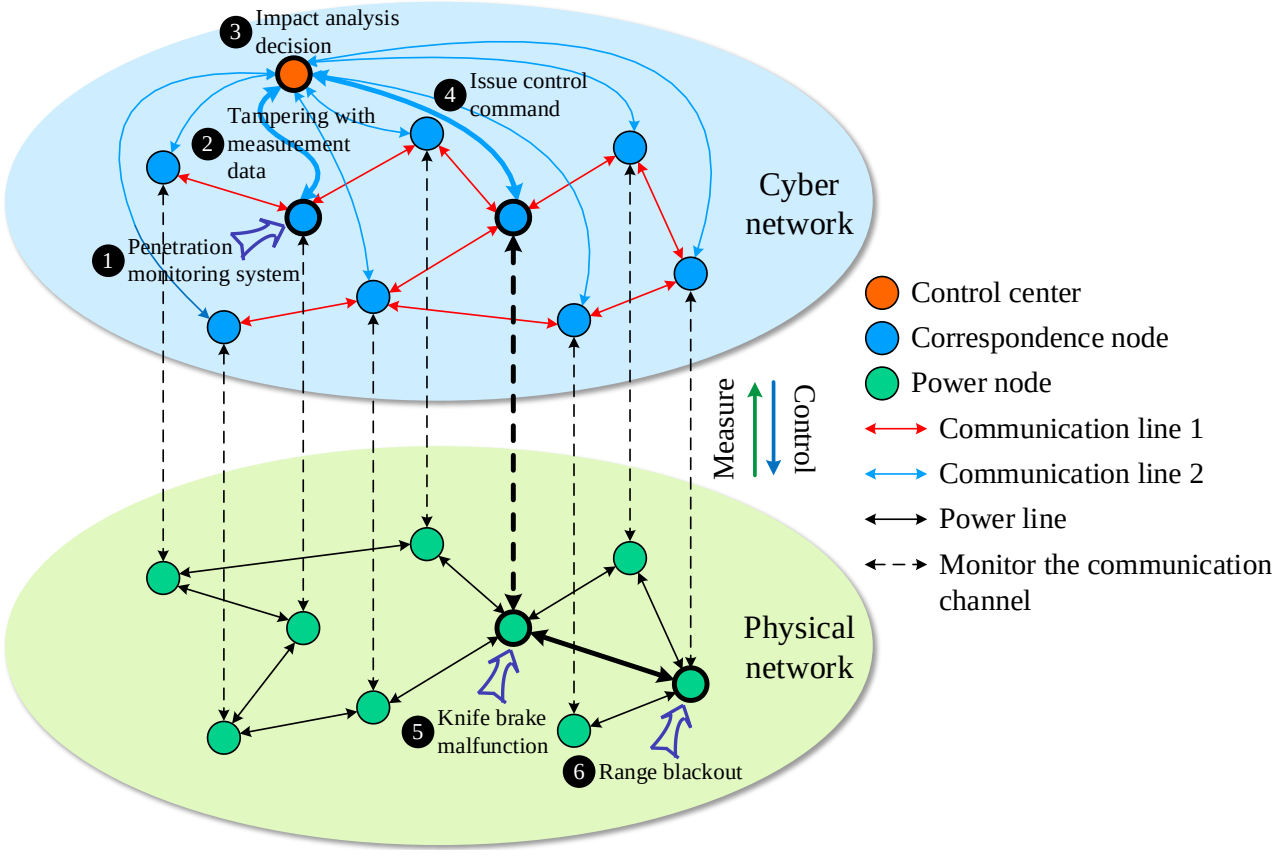

FIGURE 1. Typical power CPS FDIAs process.

Through analyzing the above process, the following questions can be drawn:

1) With certain power CPS and computer expertise, the attacker can inject false data into the intelligent measurement devices distributed in the power CPS, or directly intercept and tamper with the data packets of the SCADA system, so as to meet the conditions for avoiding the constraints of state estimation and recognition. Next, the existing bad data detection system can be successfully bypassed.

2) Power CPS information exchanges are becoming more frequent. Traditional methods of detecting FDIAs in power CPS from a knowledge-driven perspective are difficult to ensure the comprehensiveness and accuracy of detection due to the influence of the detection threshold setting. In the process of measuring data communication, state estimation system has been in a passive detection state for a long time. Once a breakthrough is made, false data will interfere with upper-level analysis and decision-making.

3) During the actual operation of power CPS, a large amount of system operation state data is generated. The existing bad data detection system has high computational complexity and poor computing power, lacks consideration of data-driven level, and neglects the advantage that artificial intelligence method can actively capture data characteristics.

### B. DETECTION ARCHITECTURE

In response to the above problems, this paper starts from the perspective of combining knowledge-driven and data-driven detection methods, and uses knowledge-driven detection methods and data-driven detection methods to perform parallel detection on measurement data. Design the active and passive hybrid detection architecture for FDIAs of power CPS, as shown in Figure 2.

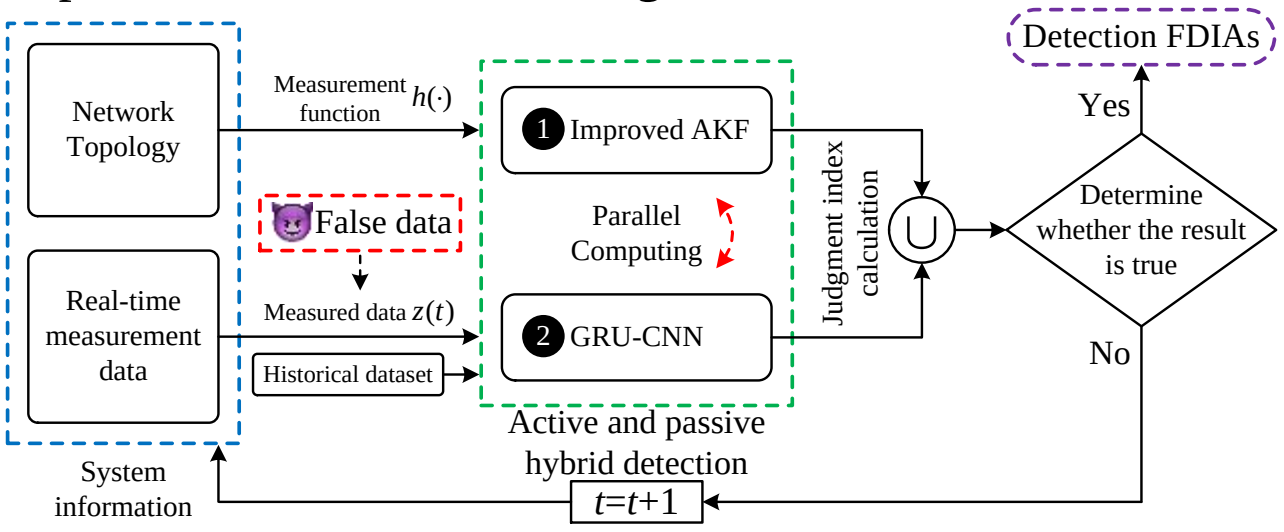

FIGURE 2. Active and passive hybrid detection architecture of power CPS FDIAs.

The network topology and system parameter matrix are obtained through the measurement function $h(\cdot)$. At the same time, the relevant parameters and data are input into the FDIAs active passive hybrid detection model by $z(t)$ obtaining the real-time measurement data at time $t$ (the GRU-CNN hybrid neural network model is trained in advance by using the historical dataset), the FDIAs passive detection method of the improved AKF and the FDIAs active detection method of the GRU-CNN hybrid neural network are calculated in parallel, and the attack judgment indicators are calculated separately. Then the respective judgment results are mixed and calculated. If the final result is true, and it is judged to be FDIAs. If it is false, then enter $t$+1 time to continue to detect the measurement data.

### C. RELATED MODELS

#### 1) STATE ESTIMATION MODEL

Taking the linear state estimation of the DC model as an example, ignoring the line resistance, the relationship between the measurement and the state can be expressed as:

$$z = Hx + e \quad (1)$$

In formula (1): $z$ is a measurement vector, including node injection active power, branch active power flow, line head power measurement value, line end power measurement value, etc. $z = [z_1, z_2, \cdots, z_m]^T$ .

$H$ is a system parameter matrix, that is, the constant Jacobian matrix, which is determined by the topology and line impedance of the power system.

$x$ is a state variable vector, including voltage phase angle of each node, etc. $x = [x_1, x_2, \cdots, x_n]^T$ .

$e$ is a measurement error, according to statistical law, it obeys a normal distribution with a mean of 0 and a variance of $\sigma^2$ .

The objective function $g(x)$ of the DC state estimation model is obtained with the traditional Weighted Least Squares (WLS) or other algorithms. The relationship between the objective function $g(x)$ and the threshold is a standard to measure whether the system has bad data. which is:

$$g(x) = (z - Hx)^T \cdot W(z - Hx) \quad (2)$$

In formula (2): $W$ is the weight value of the credibility of the measurement. The larger the weight value, the higher the measurement accuracy, and vice versa.

State estimation is to detect bad data caused by faulty sensors or topology errors, and compares the objective function $g(x)$ with a threshold $\mu$ . If $g(x) > \mu$ , it means that there is bad data. After removing the corresponding data, re-estimate the state of the system. If $g(x) \le \mu$ , it means that there is no bad data.

#### 2) STATE SPACE MODEL

As an important part of EMS, state estimation uses highly intelligent equipment such as remote terminal unit and phasor measurement unit to collect measurement information, estimate the system state, and provide the real-time operation state of the system for the dispatching center. The state variables include node voltage phase magnitude and phase angle. The research in reference [29] shows that when the power CPS is attacked or fails, its external characteristics are always expressed in the form of voltage, current or phase change. As shown in Figure 3, it is the topological structure diagram of 3-node power system [30]. The system is equipped with sensors, filters and detectors to sense, estimate and detect state parameters of FDIAs respectively.

As shown in formula (3), $S_1(t)$ is the sinusoidal form of the three-phase voltage single-point measurement signal derived from literature [31].

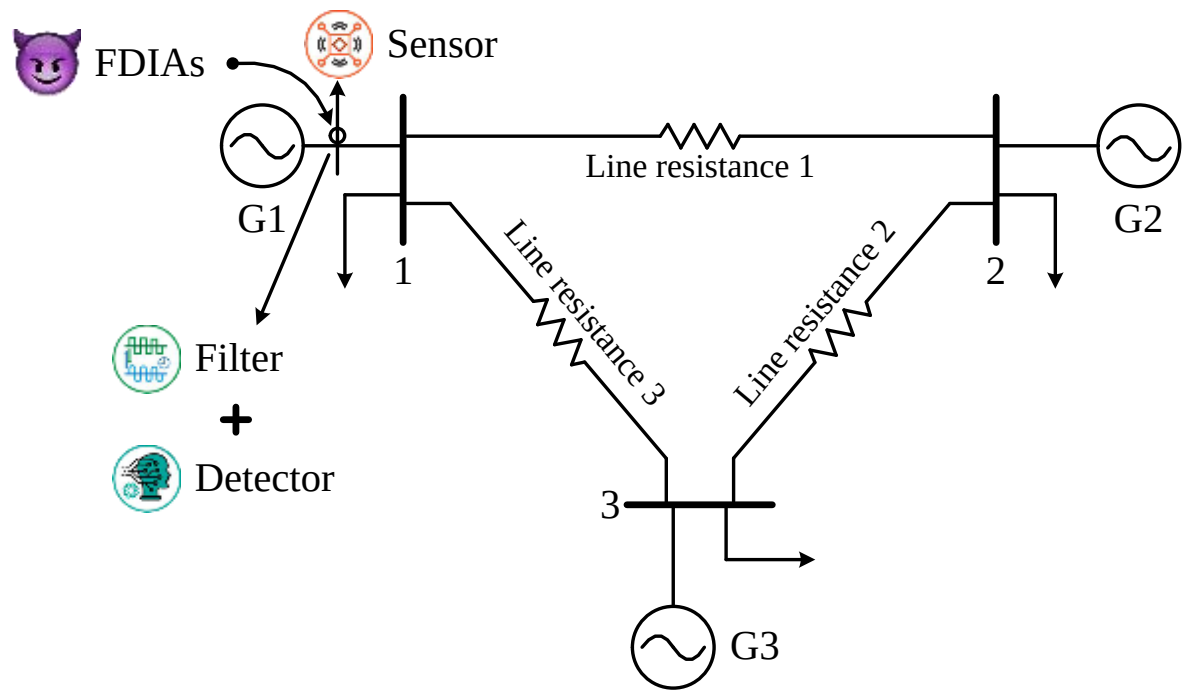

FIGURE 3. **Topological structure diagram of 3-node power system.**

$$S_1(t) = V_a sin(\omega t + \psi) = V_a sin(\omega t)cos\psi + V_a cos(\omega t)sin\psi \quad (3)$$

In formula (3): $V_a$ is the voltage magnitude, and $\omega$、$t$、$\psi$ respectively represent the angular frequency, time and phase, and $\omega$ basically does not change with the change of time $t$.

Express $V_a$ and $\psi$ as state space variables, as shown in formula (4):

$$S_1(t) = x_1 sin(\omega t) + x_2 cos(\omega t) \quad (4)$$

In formula (4): the state $x_1$ variable is equals to $V_a cos\psi$ , and $x_2$ is equals to $V_a sin\psi$ .

As time $t$ changes, formula (5) shows the process noise equation of state.

$$x(t+1) = \begin{pmatrix} 1 & 0 \\ 0 & 1 \end{pmatrix} x(t) + \eta(t) \quad (5)$$

In formula (5): $\eta(t)$ is the process noise sequence. The system state $x(t)$ is equals to $[V_a cos\psi \ V_a sin\psi]^T$.

As shown in formula (6), it is the actual voltage signal of the current state.

$$z(t) = (cos\omega t - sin\omega t)\begin{bmatrix} x_1(t) \\ x_2(t) \end{bmatrix} + \zeta(t) \quad (6)$$

In formula (6): $z(t)$ is the measurement vector at time $t$. $\zeta(t)$ is the measurement noise sequence at time $t$.

To further organize the formula (5)-(6) representing the state space model, it can be expressed as:

$$x(t+1) = Bx(t) + \eta(t) \quad (7)$$

$$z(t) = Hx(t) + \zeta(t) \quad (8)$$

In formula (7)-(8): system matrix $B = \begin{pmatrix} 1 & 0 \\ 0 & 1 \end{pmatrix}$ , and observation matrix $H$ is equals to $cos\,\omega\, t$-$sin\,\omega\, t$. Assuming that the sum $\eta(t)$ and $\zeta(t)$ are the mean of 0, the standard deviation $\sigma$ is Gaussian white noise, and the two are uncorrelated, then the formula (9) is satisfied:

$$\begin{cases} G\{\eta_t\} = G\{\zeta_t\} = 0 \\ G\{\eta_t \zeta_i^T\} = 0 \\ G\{\eta_t \zeta_i^T\} = \begin{cases} \hat{M}_t, i = t \\ 0, i \neq t \end{cases} \\ G\{\eta_t \zeta_i^T\} = \begin{cases} \hat{N}_t, i = t \\ 0, i \neq t \end{cases} \end{cases} \tag{9}$$

In formula (9): $\hat{M}_t$ is the covariance matrix of process noise, and $\hat{N}_t$ is the covariance matrix of measurement noise.

#### 3) FDIAS MODEL

According to the literature [32], the FDIAs model can be expressed as:

$$z_{ac}(t) = Hx_{ac}(t) + \zeta(t) + \Phi y_{ac}(t) \tag{10}$$

In formula (10): $z_{ac}(t)$ and $x_{ac}(t)$ represent the amount of false data that has been injected. Sensor selection matrix $\Phi$ $=diag(\delta_1, \delta_2, \cdots, \delta_n)$, if the $i$-th sensor is attacked, $\delta_i=1$. $y_{ac}(t)$ is a carefully designed attack sequence for the attacker. In the case of random attacks with false data, $y_{ac}(t)$ is an arbitrary sinusoidal attack signal.

Assuming that $ac$ is a non-zero attack vector and injects into the measurement data $z$, then $z_{fd}=z+ac$ represents the measurement data result after $z$ is attacked. The deviation vector of the estimated state variables before and after the attack is expressed by $d$, then the estimated system state vector after the attack can be expressed as $\hat{x}_{fd} = \hat{x} + d$. At this time, $e_{ac}$ is used to represent the residual after being attacked.

$$\begin{aligned} e_{ac} &= \|(z+ac) - H(\hat{x}+d)\| = \|(z - H\hat{x}) + (ac - Hd)\| \\ &\leq \|z - H\hat{x}\| + \|ac - Hd\| \end{aligned} \tag{11}$$

With knowing the network parameters and topology of power CPS, the attacker manipulates the measuring device to inject false data into the communication transmission value of the measuring device. When $ac=Hd$, the attacker can successfully pass the system data detection process, so as to manipulate the state estimation results to attack power CPS.

In the above, relevant models such as state estimation, state space and FDIAs are established. In view of the shortcomings of the traditional knowledge driven detection method proposed in the above problem description, and the lack of consideration on the data-driven level in the existing bad data detection system. The improved AKF algorithm and GRU-CNN hybrid neural network will be used to realize the effective detection of FDIAs.

## III. PASSIVE DETECTION METHOD OF FDIAS IN THE POWER CPS BASED ON IMPROVED AKF

First, analyze the AKF algorithm, then improve the AKF algorithm's shortcomings, and finally propose a passive detection method for FDIAs with the improved AKF algorithm.

### A. AKF ALGORITHM ANALYSIS

#### 1) AKF ALGORITHM PRINCIPLE

The state space model is established for the three-node power system above, as shown in formula (7)-(8), the AKF algorithm used is as follows.

Residual $e_t$:

$$e_t = \hat{z}_t - H_t \hat{x}_{t,t-1} - \hat{s}_t \tag{12}$$

In formula (12): $\hat{z}_t$ is the estimated value of the measurement at time $t$. $H_t$ is the measurement matrix. $\hat{x}_{t,t-1}$ is the pre-estimated state value for the previous moment. $\hat{s}_t$ is the mean value of the measured noise. ˆ represents the estimated value of the corresponding quantity.

Mean square error $Q_t$ and state prediction $\hat{x}_{t,t-1}$ for the next moment:

$$Q_{t,t-1} = BQ_{t-1}B^T + U_{t-1}\hat{M}_{t-1}U_{t-1}^T \tag{13}$$

$$\hat{x}_{t,t-1} = B\hat{x}_{t-1} + U_{t-1}\hat{p}_{t-1} \tag{14}$$

In formula (13)-(14): $U_{t-1}$ is the system noise driving matrix. $\hat{p}_{t-1}$ is the mean value of process noise.

Update amount of Kalman measurement value:

$$E_t = Q_{t,t-1}H_t^T(H_t Q_{t,t-1} H_t^T + \hat{N}_t)^{-1} \tag{15}$$

$$\hat{x}_t = \hat{x}_{t,t-1} + E_t e_t \tag{16}$$

$$Q_t = (I - E_t H_t) Q_{t,t-1} \tag{17}$$

In formula (15)-(17): $E_t$ is Kalman gain. $I$ is the identity matrix.

The adaptive estimation algorithms are:

$$\hat{p}_t = (1-c_t)\hat{p}_{t-1} + c_t[\hat{x}_t - B\hat{x}_{t-1}] \tag{18}$$

$$\hat{M}_t = (1-c_t)\hat{M}_{t-1} + c_t[E_t e_t e_t^T E_t^T + Q_t - BQ_{t-1}B^T] \tag{19}$$

$$s_t = (1-c_t)\hat{s}_{t-1} + c_t[\hat{z}_t - Hx_{t,t-1}] \tag{20}$$

$$\hat{N}_t = (1-c_t)\hat{N}_{t-1} + c_t[e_t e_t^T - HQ_{t,t-1}H^T] \tag{21}$$

$$c_t = \frac{1-g}{1-g^{t+1}}, \quad 0 < g < 1 \tag{22}$$

In formula (18)-(22): $c_t$ is the weighting coefficient, $g$ is the forgetting factor, usually the value is 0.95 ~ 0.99 [33].

#### 2) AKF ALGORITHM DEFICIENCY

Although the AKF algorithm has certain advantages in the detection of FDIAs in power CPS, it also has shortcomings, mainly as follows:

(1) There is a term to subtract the mean square error $Q_{t,t-1}$ in formula (21). When applied to higher-order systems, as the $\hat{N}_t$ and $\hat{M}_t$ continuously updated, the filtering is prone to divergence. At this time, the positive definite diagonal matrix $\hat{N}_t$ and the semi-positive definite diagonal matrix $\hat{M}_t$ cannot guarantee the non-negativity of diagonal elements.

(2) Power CPS requires high real-time data processing. Considering that $\hat{N}_t$ and $\hat{M}_t$ need to be updated continuously, due to the limitation of computing power, it is difficult to guarantee the stability of filtering and the speed of attack detection.

Therefore, the traditional AKF algorithm is difficult to meet the actual needs of power CPS FDIAs detection, and it needs to be further improved.

### B. AKF ALGORITHM IMPROVEMENT

It can be seen from the literature [34] that in the process of power CPS FDIAs detection, the measurement noise is known and the process noise is unstable, so only needs to be estimated $\hat{M}_t$ and $\hat{N}_t$ can be set as a fixed value. In view of the above situation, the mean value of the measurement noise $\hat{s}$ in the residual is set to 0, that is, formula (12) is updated to:

$$e_t = \hat{z}_t - H_t \hat{x}_{t,t-1} \tag{23}$$

Formula (12) is updated to formula (23), and filter calculation formula (14) and (16) are also updated accordingly. Since the process noise covariance matrix $\hat{M}_t$ has a term that subtracts the mean square error, the $Q_t$ and $Q_{t,t-1}$ in the formula (19) are also simplified to 0. The system matrix $B$ and the state quantity $\hat{x}_t$ at the adjacent moments before and after determine the update of $\hat{p}_t$ in the adaptive estimation algorithm, and the adaptive estimation algorithm is updated as:

$$\hat{p}_t = (1-c_t)\hat{p}_{t-1} + c_t[\hat{x}_t - B\hat{x}_{t-1}] \tag{24}$$

$$\hat{M}_t = (1-c_t)\hat{M}_{t-1} + c_t E_t e_t e_t^T E_t^T \tag{25}$$

The mean square error term in the noise covariance matrix $\hat{M}_t$ is deleted, and other retained residual terms can still ensure the accuracy and stability of the filtering when $\hat{M}_t$ is continuously updated. The improved AKF algorithm avoids the filtering divergence problem caused by the $\hat{M}_t$ non-negative positive definite matrix. At the same time, the simplification of the algorithm structure also improves the calculation speed and improves the timeliness of FDIAs detection.

### C. PASSIVE DETECTION OF FDIAS

This paper is based on the Euclidean distance detection method and residual error detection method to detect and analyze FDIAs in power CPS.

#### 1) EUCLIDEAN DISTANCE DETECTION METHOD

In the power CPS, the attacker's carefully designed FDIAs sequence can completely invalidate the traditional bad data detection mechanism, thereby affecting the state estimation. In this paper, the Euclidean distance detection method in the similarity algorithm is used to calculate the difference between the observed value and the estimated value [31]. To achieve the purpose of detecting FDIAs.

In order to reduce the false alarm rate caused by noise, when applying the Euclidean distance detection method, the threshold is set to 3 times the standard deviation, that is, $3\sigma$, and the false alarm rate can be reduced to 0.27% through experiments [35].

If the deviation $d[S(t), S_a(t)]$ exceeds the set threshold, it will be judged as being attacked at this time, that is, the system has suffered a FDIAs.

$$d[S(t), S_a(t)] = \sqrt{\left[H\hat{x}_a(t) - Hx(t)\right]^2} \tag{26}$$

In formula (26): $\hat{x}_a(t)$ and $S_a(t)$ represent the amount of false data that has been injected. $S(t)$ and $S_a(t)$ represent the observed value and estimated value of the voltage signal magnitude respectively.

#### 2) RESIDUAL DETECTION METHOD

Considering the situation that $ac=Hd$ in formula (11), $\chi^2$ detection based on residual error can no longer successfully detect FDIAs. At this time, consider the residual detection method based on voltage state analysis:

$$r(x,\hat{x}) = \frac{\| x(t) - \hat{x}(t) \|}{\| x(t) \| \cdot \| \hat{x}(t) \|} \tag{27}$$

In formula (27): when there is no FDIAs, the observed value is the same as the actual value, that is, $\lim_{t\to\infty} r(x,\hat{x}) = 0$. otherwise, when $\lim_{t\to\infty} \| x_a(t) - x(t) \| \geq \alpha(\alpha > 0)$, $\lim_{t\to\infty} r(x,\hat{x}) > \rho(\rho > 0)$. Considering the influence of noise, the threshold of residual detection is also set to $3\sigma$. When $r(x,\hat{x}) \geq 3\sigma$, it will be judged as being attacked, otherwise, it proves that the system is not attacked.

Through the above analysis, the state estimation value is obtained based on the improved AKF algorithm. Both the Euclidean distance detection method and the residual detection method can detect FDIAs from the system state estimation value.

## IV. ACTIVE DETECTION METHOD OF FDIAS FOR POWER CPS BASED ON GRU-CNN HYBRID NEURAL NETWORK

### A. GRU AND CNN THEORETICAL ANALYSIS

With the advancement of smart grid and energy internet strategies, high-dimensional multi-label samples such as node structure, connection characteristics, content characteristics, and external characteristics based on time and space have been introduced. The method of selecting local features based on prior knowledge to improve detection accuracy is no longer suitable for attack detection in power CPS scenarios, deep learning is an important branch of machine learning. The real-time massive data generated by power CPS is more in line with the characteristics of deep learning algorithms. It can improve the ability to extract data features by constructing deep and complex neural network architectures.

GRU is a special kind of Recurrent Neural Network (RNN). Compared with Long Short-Term Memory (LSTM), GRU has more reasonable structure, parameters and convergence. It optimizes the calculation method of the hidden state in the recurrent neural network, in which the reset gate can be used to discard historical information irrelevant to the prediction, and the update gate can control how the hidden state is updated by the candidate hidden state. In this paper, GRU is used to extract the data characteristics of power CPS, and the concepts of reset gate $R_t$ and update gate $Z_t$ are introduced. The specific formula are as follows:

$$R_t = \sigma(X_t W_{xr} + H_{t-1} W_{hr} + b_r) \tag{28}$$

$$Z_t = \sigma(X_t W_{xz} + H_{t-1} W_{hz} + b_z) \tag{29}$$

$$\tilde{H}_t = tanh(X_t W_{xh} + (R_t \odot H_{t-1}) W_{hh} + b_h) \tag{30}$$

$$H_t = Z_t \odot H_{t-1} + (1 - Z_t) \odot \tilde{H}_t \tag{31}$$

In formula (28)-(31): $t$ is the time step. $W_{xr}$, $W_{hr}$, $W_{xz}$, $W_{hz}$, $W_{xh}$, $W_{hh}$ are the weight parameters. $b_r$, $b_z$, $b_h$ are the deviation parameters, all obtained by training. $\sigma()$ is the Sigmoid activation function, which transforms the element range to [0, 1]. $\odot$ means do element multiplication. The input of $R_t$ and $Z_t$ is the current time input $X_t$ and the previous hidden state $H_{t-1}$. $H_t$ is the hidden state. $\tilde{H}_t$ is the candidate hidden state.

As shown in Figure 4, it is a schematic diagram of the CNN process. The basic unit in the convolutional layer is the convolution layer followed by the maximum pooling layer, which is classified by the fully connected layer and the output layer after multiple rounds of processing [36]. The convolution layer is used to identify the spatial pattern in the real-time feature data matrix, and the pooling layer is used to reduce the sensitivity of the convolution layer to position. The fully connected layer and the output layer function as feature classifiers.

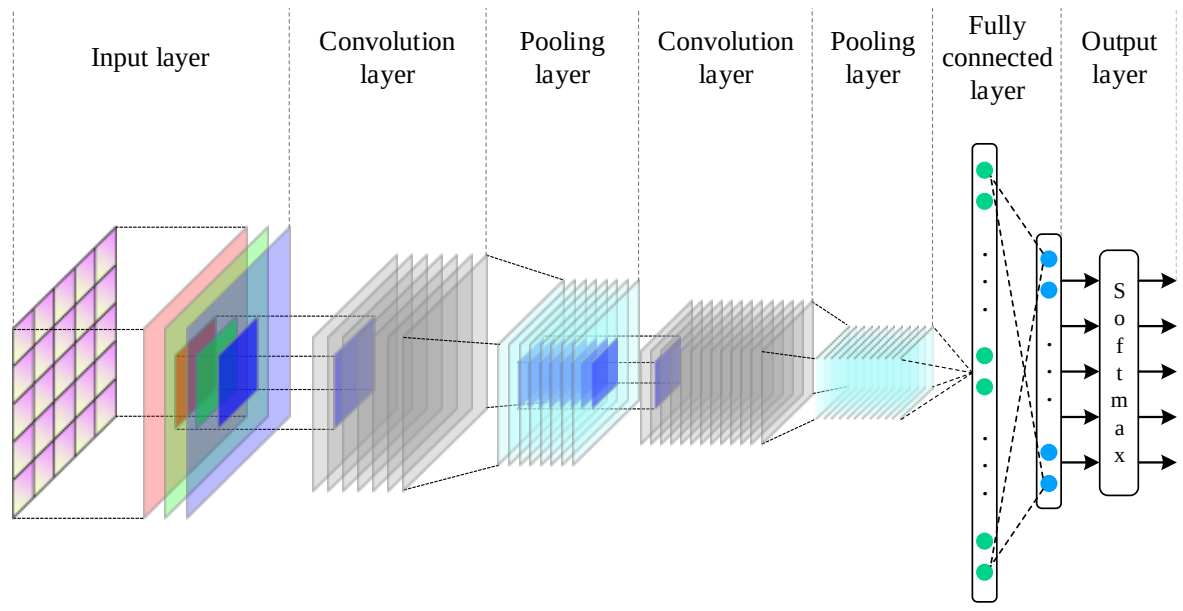

FIGURE 4. **Schematic diagram of the CNN process.**

The convolution process and the label probability distribution formula of the fully connected layer are as follows [37]:

$$A_{i,j} = f(\sum_{m=0}\sum_{n=0} w_{m,n} H_{i+m,j+n} + b) \tag{32}$$

$$S(\tilde{y} = p \mid x) softmax(y_p) = \frac{exp(y_p)}{\sum_{p=1}^{q} y_p} \tag{33}$$

In formula (32)-(33): $A_{i,j}$ is the output of the convolution layer. $w_{m,n}$ is the element in the $m$th row and $n$th column of the convolution kernel matrix. $H_{i+m,j+n}$ is the $(i+m)$th row and $(j+n)$th column element of the convolution input matrix. $b$ is bias. $f$ is the Relu activation function. $S$ is the posterior probability. $x$ is the input of the fully connected layer. And $y_p$ is the output category, the output of the fully connected layer containing $q$ elements.

Combining the strong memory of the recurrent neural network and the strong feature expression ability of the convolutional neural network in deep learning, the GRU output $H_t$ is used as the data input of the CNN, and the CNN network performs multi-dimensional features through 2 convolutions and poolings in turn extraction, and finally realize whether the power CPS is subject to FDIAs detection.

### *B. STRUCTURE DESIGN OF GRU-CNN HYBRID NEURAL NETWORK*

Before performing FDIAs detection, it is necessary to complete the design of the GRU-CNN hybrid neural network structure. The general structure of the network is shown in Figure 5.

The GRU-CNN hybrid neural network structure designed in this paper fully considers the characteristics of power CPS measurement data. The A, B, C three-phase voltage (current) phase angle, phase magnitude, positive sequence, negative sequence, zero sequence voltage (current) phase angle, phase magnitude, etc. are preprocessed as the input of the hybrid neural network. Its network structure is mainly composed of 6 parts, namely:

1) Sequence feature extraction layer. The sequence feature extraction layer adds 100 GRU structures to process and extract the time features of the input data. Its single GRU structure consists of update gates and reset gates.

2) Input layer. The input layer is the input of the entire CNN. In the process of data processing, each sample data is input in a two-dimensional form.

3) Convolution layer. The convolution layer is the most important part of the CNN, used to extract the local features of the input neuron data. Each convolution layer is composed of multiple feature matrices. Each feature matrix is a plane. The corresponding convolution kernels on the same plane are the same. They have the characteristics of rotation, displacement invariance and weight sharing, as well as parallel learning. The number of free parameters is greatly reduced. Different planes correspond to different convolution kernels, and multiple convolution kernels make feature extraction more sufficient. For a two-dimensional input feature matrix $Z$ and a learnable convolution kernel $K$ for two-dimensional convolution, the convolution operation

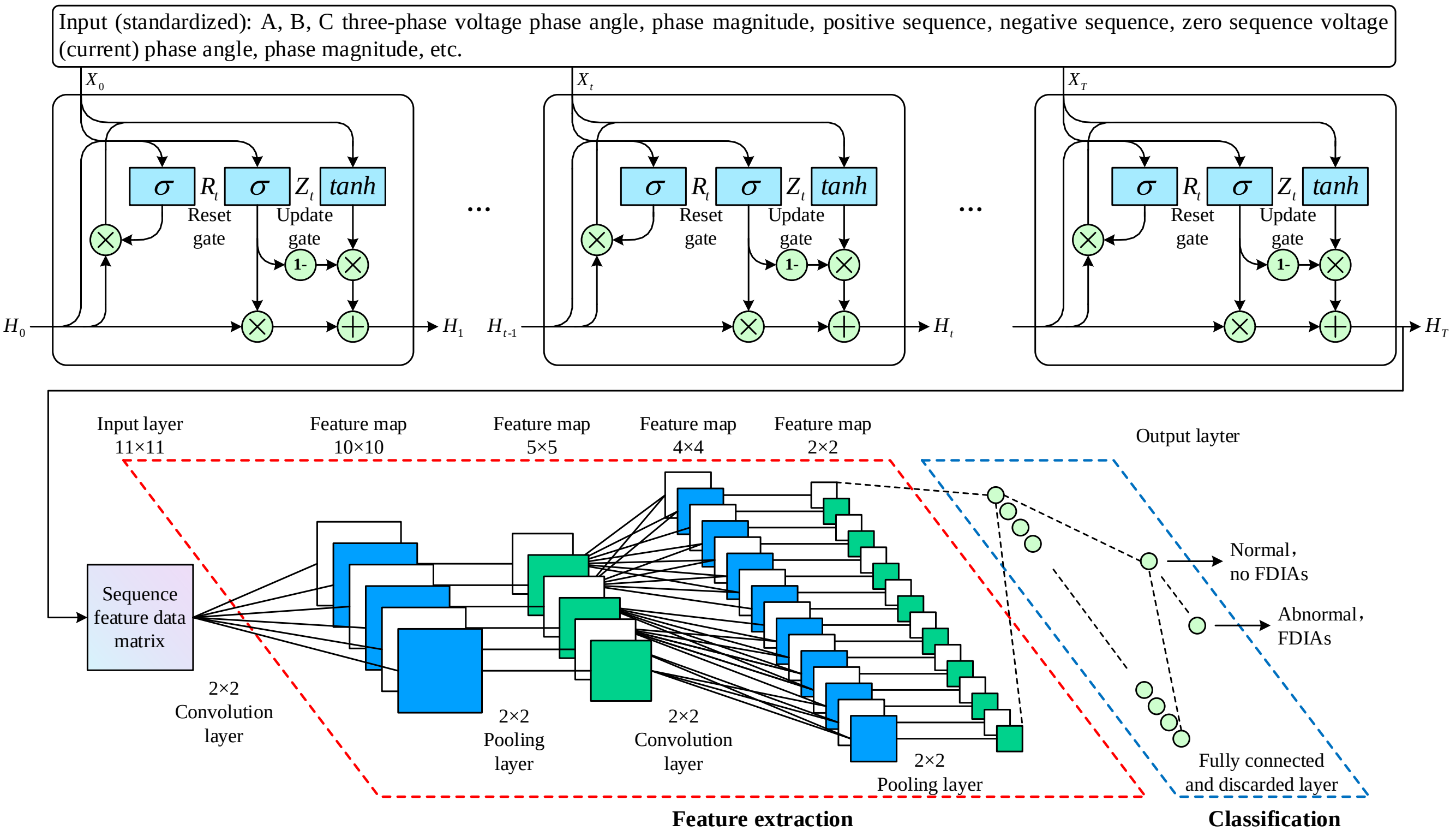

**FIGURE 5. Schematic diagram of the GRU-CNN hybrid neural network structure.**

can be expressed as:

$$S(i,j)=(ZK)(i,j)=\sum_{m=1}^{i}\sum_{n=1}^{j}Z(m,n)K(i-m,j-n) \quad (34)$$

In formula (34): $i$ and $j$ are the positions of the horizontal and vertical axes of the blocks corresponding to the previous layer of neural network. $m$ and $n$ are the positions of the horizontal and vertical axes of the blocks corresponding to the convolution layer.

4) Pooling layer. The pooling layer is used to scale and map the data of the upper layer to reduce the data dimension. The extracted features are scale-invariant and can also prevent overfitting. Generally use mean pooling (Pooling), such as downsampling [38].

5) Fully connected and discarded layer. After multiple rounds of convolution layer and pooling layer processing, the input information has been abstracted into features with higher information content, and the classification result is given by the fully connected layer at the end of the CNN. At the same time, in order to solve the problem of time-consuming and overfitting, this paper adds a discard layer to weaken the joint adaptability between neurons. In the training phase, some elements are randomly selected with a weight of 0 and discarded from the network to enhance the generalization ability.

6) Output layer. After the fully connected and discarded layer, CNN mainly uses the Softmax classifier to classify and output the results.

### C. ACTIVE DETECTION OF FDIAS

#### 1) CHARACTERISTIC DATA EQUALIZATION PROCESSING

In the long-term operation of power CPS, the probability of FDIAs is low. It is difficult for the measurement system to obtain sufficient data to characterize the state of the power grid when the attack occurs. The low number of attack samples will cause serious imbalance of data, and the false alarm rate of attack detection model is too high. In this paper, CKS (Central-Kmeans-Smote) oversampling algorithm is used to generate pseudo samples which are highly similar to real attack samples and add them to the original sample set, so that the number of attack samples and normal samples tends to be balanced, so as to realize the equalization of attack data [39]. The process of feature data equalization is shown in Figure 6.

#### 2) CHARACTERISTIC DATA STANDARDIZATION PROCESSING

In order to ensure the certainty of the output, this paper uses the Z-score method to standardize the original time series data, and replace the missing features with the mean value of the feature. The standardized data $Z$ is used as the input of the GRU model.

$$Z=\frac{x-\alpha}{\sigma} \quad (35)$$

In formula (35): $\alpha$ is the mean value of the feature on the entire dataset. $\sigma$ is the standard deviation.

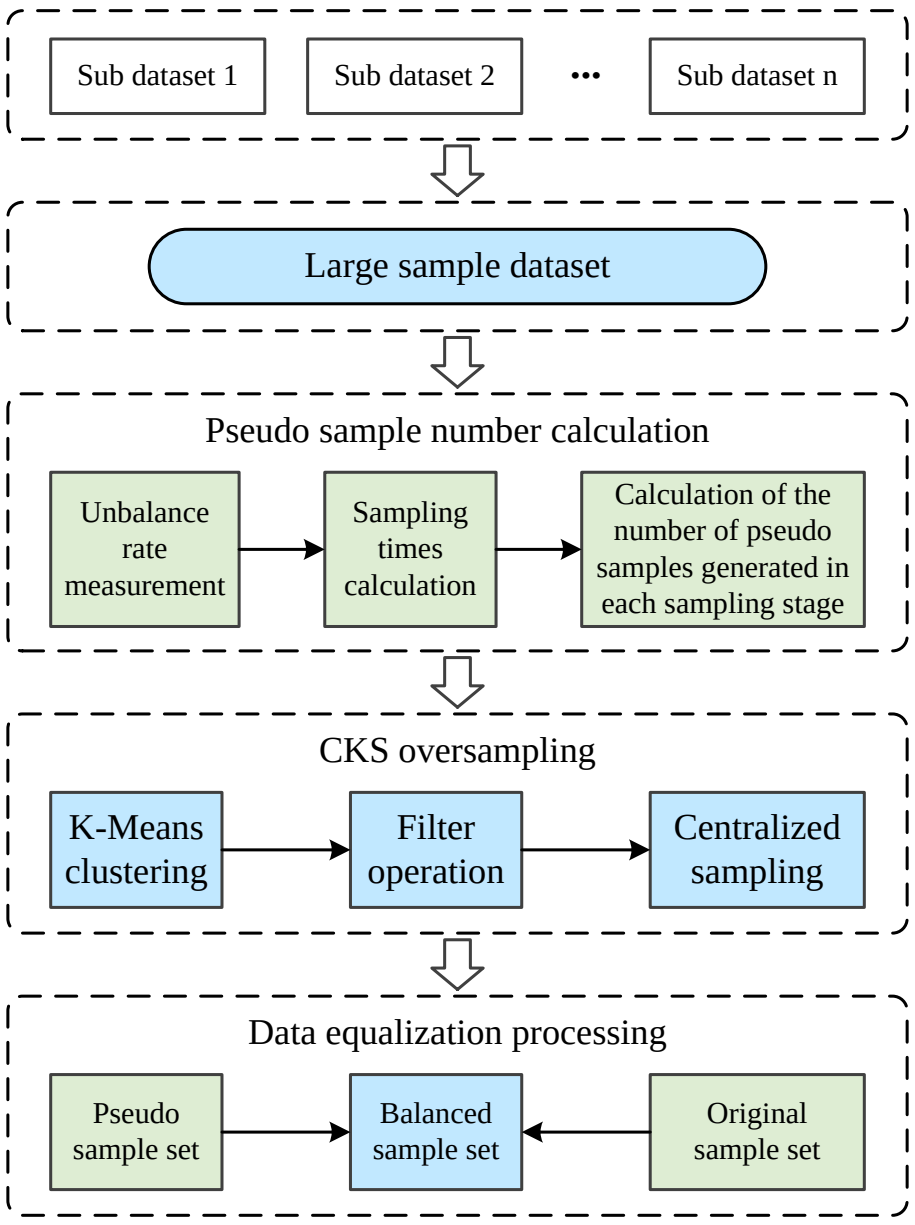

**FIGURE 6. Feature data equalization process.**

### 3) GRU-CNN HYBRID NEURAL NETWORK MODEL TRAINING AND DETECTION

After training the GRU-CNN hybrid neural network structure according to the above description, it can be used for FDIAs detection. In this paper, Adam algorithm is used to optimize training efficiency [40], combined with momentum variables and RMSProp algorithm, to set the corresponding learning rate for the gradient.

FDIAs detection steps with GRU-CNN hybrid neural network are as follows:

Step 1: Preprocess the measurement dataset $\{z_i\}$, and process the $n$ measured value vectors $z$ into a $n \times m$ matrix $Z$, the expression is:

$$Z = \begin{bmatrix} z_{11} & z_{12} & \cdots & z_{1m} \\ z_{21} & z_{22} & \cdots & z_{2m} \\ \cdots & \cdots & \ddots & \cdots \\ z_{n1} & z_{n2} & \cdots & z_{nm} \end{bmatrix} \tag{36}$$

The matrix Z is used as the input of the GRU-CNN hybrid neural network.

Step 2: Input the preprocessed input layer data into 100 GRU structures. Through the update gate and reset gate, the update gate is used to control the extent to which the state information at the previous moment is brought into the current state. The larger the value of the update gate, the more state information from the previous moment is brought in. The reset gate is used to control the degree of ignoring the state information at the previous moment. The smaller the reset gate value, the more ignored.

Step 3: After passing through the GRU structure, input the obtained result to the convolution layer. The convolutiona layer is divided into two parts: a feature extraction layer and a feature mapping layer. The input of each neuron in the feature extraction layer is connected to the local receiving area of the previous layer, and the local feature is extracted (once the local feature is extracted, the positional relationship between it and other features is also determined). Each computing layer of the network in the feature mapping layer is composed of multiple feature mappings, each feature mapping is a plane, and the weights of all neurons on the plane are equal. The feature mapping structure adopts the Sigmoid function with a small influence function core as the activation function of the convolution network, so that the feature mapping has displacement invariance.

Step 4: After the convolution layer is processed, the result is input to the pooling layer, the input data is divided into non-overlapping regions according to the window size, and then the elements in each region are aggregated. According to the size of the data volume, the appropriate window size is selected for the pooling operation, so as to realize the dimensionality reduction of the features.

Step 5: Input the obtained result into the final fully connected and discard layer, and use the Softmax classifier to classify and output the result.

## V. ACTIVE AND PASSIVE HYBRID DETECTION METHOD FOR POWER CPS FDIAS

According to the deficiencies of the knowledge-driven and data-driven detection methods proposed in the second section relative to each other, this paper proposes an active and passive hybrid detection method, while introducing the third section with the passive detection method of FDIAs based on improved AKF and the fourth section with active detection method of FDIAs based on GRU-CNN hybrid neural network to deal with FDIAs, the results obtained by the two detection methods are mixed and fixed. The active and passive hybrid detection method is as follows:

Suppose at time $t$, the residual value obtained by the passive detection method of FDIAs based on improved AKF is $\hat{r}_t^N$, the residual detection threshold is set to $3\sigma$, and the boolean value obtained by the active detection method of FDIAs based on GRU-CNN hybrid neural network is $\hat{r}_t^{GC}$. The residual value $\hat{r}_t^N$ and the boolean value $\hat{r}_t^{GC}$ obtained by the two detection methods are mixed and fixed to obtain the active and passive hybrid detection boolean value $\hat{r}_t$:

$$\hat{r}_t = f(\hat{r}_t^N, 3\sigma) \cup \hat{r}_t^{GC} \tag{37}$$

$$\hat{r}_t = \begin{cases} 0 & \text{no FDIAs} \\ 1 & \text{FDIAs} \end{cases} \tag{38}$$

In formula (37): $f$ is the true value function of whether the residual value $\hat{r}_t^N$ is greater than the residual detection threshold $3\sigma$.

The active and passive hybrid detection process of FDIAs is shown in Figure 7.

Improved AKF's FDIAs passive detection method is difficult to actively capture data features, and the detection accuracy is greatly affected by the threshold setting. The FDIAs active detection method of the GRU-CNN hybrid neural network is difficult to explain from the mechanism and convincing people. When new scenes appear, it takes time to complete the update, and the timeliness is poor. Improved AKF and GRU-CNN to complete parallel operations independently, and the two do not affect each other. After that, the active and passive fixed operations are mixed and the results of the two operations are taken. Therefore, the active and passive hybrid detection method proposed in this paper can combine the advantages of improved AKF state estimation and online calculation with good timeliness, and the advantages of active capture data based on GRU-CNN to make full use of historical data and not be affected by FDIAs. Its advantages make up for the shortcomings of the other party, can get better detection results and can cope with FDIAs.

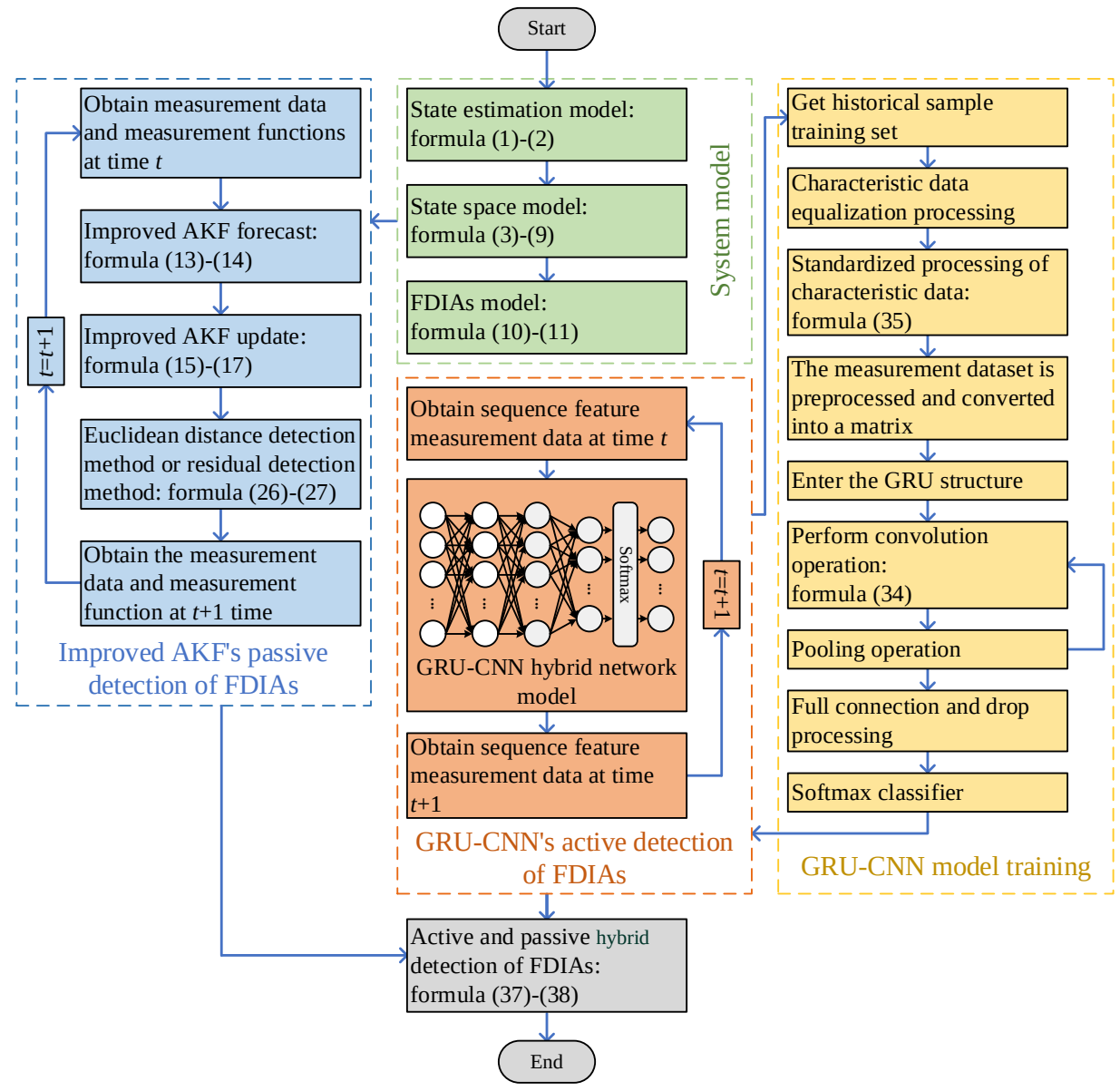

**FIGURE 7.** **Active and passive hybrid detection process of FDIAs.**

## VI. EXAMPLE ANALYSIS

The experimental environment in this paper uses the simulation system shown in Figure 8. The behavior of FDIAs and the generation of operating data are all completed in this environment. FDIAs are accurately detected by the active and passive hybrid detection method mentioned in this paper, and then the effectiveness of the method is verified.

The power CPS simulation system shown in Figure 8 consists of a three-layer structure of perceptual executive layer, communication layer and decision control layer. The perceptual executive layer includes two generators, G1 and G2. BR1 to BR4 are circuit breakers, which are monitored by R1 to R4 Intelligent Electronic Devices (IED), which can open or close the corresponding circuit breakers. There are two lines in the power grid, the first line L1 extends from BR1 to BR2, and the second line L2 extends from BR3 to BR4. Each IED uses a distance protection scheme, when a fault is detected, since they do not have internal verification to detect anomalies, no matter whether the fault is real or valid, it will trigger the circuit breaker. The communication layer includes substation switch, which is connected to IEDs, PDC, and router to complete the uplink (information collection) and downlink (optimized control) communication processes. Uplink communication requires a PDC to summarize the phasor data collected by IED and upload it to OpenPDC. The decision control layer includes the Snort intrusion detection system, system log, control panel and OpenPDC. Snort performs intrusion detection on the communication layer substation switch, and the control panel displays the real-time time series data processed by OpenPDC. Analyze and make decisions with the operating status, and the system log records communications, intrusion detection, and system operating status.

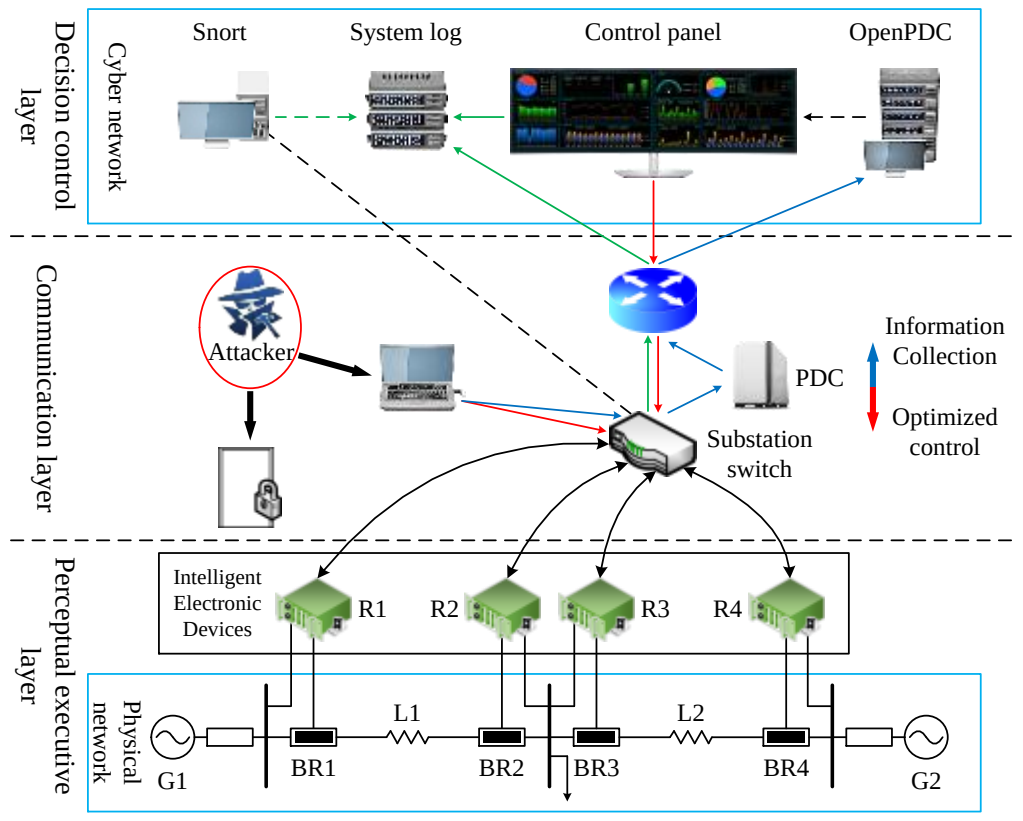

**FIGURE 8.** **Schematic diagram of power CPS simulation system structure.**

In order to verify the timeliness and accuracy of the FDIAs active-passive hybrid detection method proposed in this paper, a case study is carried out from the following three aspects: 1) Simulation analysis of FDIAs passive detection based on improved AKF. 2) Simulation analysis of FDIAs active detection based on GRU-CNN. 3) Simulation analysis of active and passive hybrid detection.

### A. SIMULATION ANALYSIS OF FDIAS PASSIVE DETECTION BASED ON IMPROVED AKF

As known from above, the mean square error term in the noise covariance matrix $\hat{M}_t$ was deleted. The improved AKF algorithm avoids the filtering divergence problem caused by the non-negative positive definite matrix of $\hat{M}_t$, and the simplification of the algorithm structure also improves the calculation speed. In order to verify the timeliness of the improved AKF FDIAs passive detection method, the voltage phase magnitude on BR3 in the power CPS simulation system was selected, the amount of false data injection was 5% of the original measurement, and 3254 samples of R3 data were taken continuously, and the false data were injected at intervals.

Figure 9 shows the comparison of FDIAs passive detection under different filtering algorithms. It can be seen from Figure 9 that when there was false data in the R3 measurement, the improved AKF algorithm was compared with the unimproved AKF algorithm, whether using Euclidean detection method or residual detection method, the former had a faster response speed, and the absolute value $|\hat{x}_a(t+1)-\hat{x}(t)|$ of the difference between the time $t$+1 after being attacked and the previous time $t$ was significantly higher than that of no attack. The improved AKF Euclidean detection method reached the threshold at 2278, and FDIAs was detected. Compared with the AKF Euclidean detection method, the time was shortened by 7. The improved AKF residual detection method was even after the attack started, the threshold was exceeded in a very short time, and the detection rate was significantly increased by 97.3%, which verified the rationality of the improved AKF algorithm. In view of the fact that the simulation system shown in Figure 8 ignores the anomaly detection of the attacked IED measurement, the improved AKF passive detection method of FDIAs proposed in this paper can effectively detect the false data on the R3 measurement, effectively solving the original algorithm filtering divergence and calculation speed restricted issues.

In addition, in order to verify the accuracy of the improved AKF FDIAs passive detection method, a certain time $t$ was selected to attack the R1-R4 of the power CPS simulation system by constructing a false data attack vector, using the weighted least squares method and the improvement proposed in this paper, respectively. The AKF algorithm compares the estimated values of the three-phase voltage phase magnitude and voltage phase angle of A, B, and C under different conditions of non-attack and attacked. The comparison of the estimated results is shown in Figure 10.

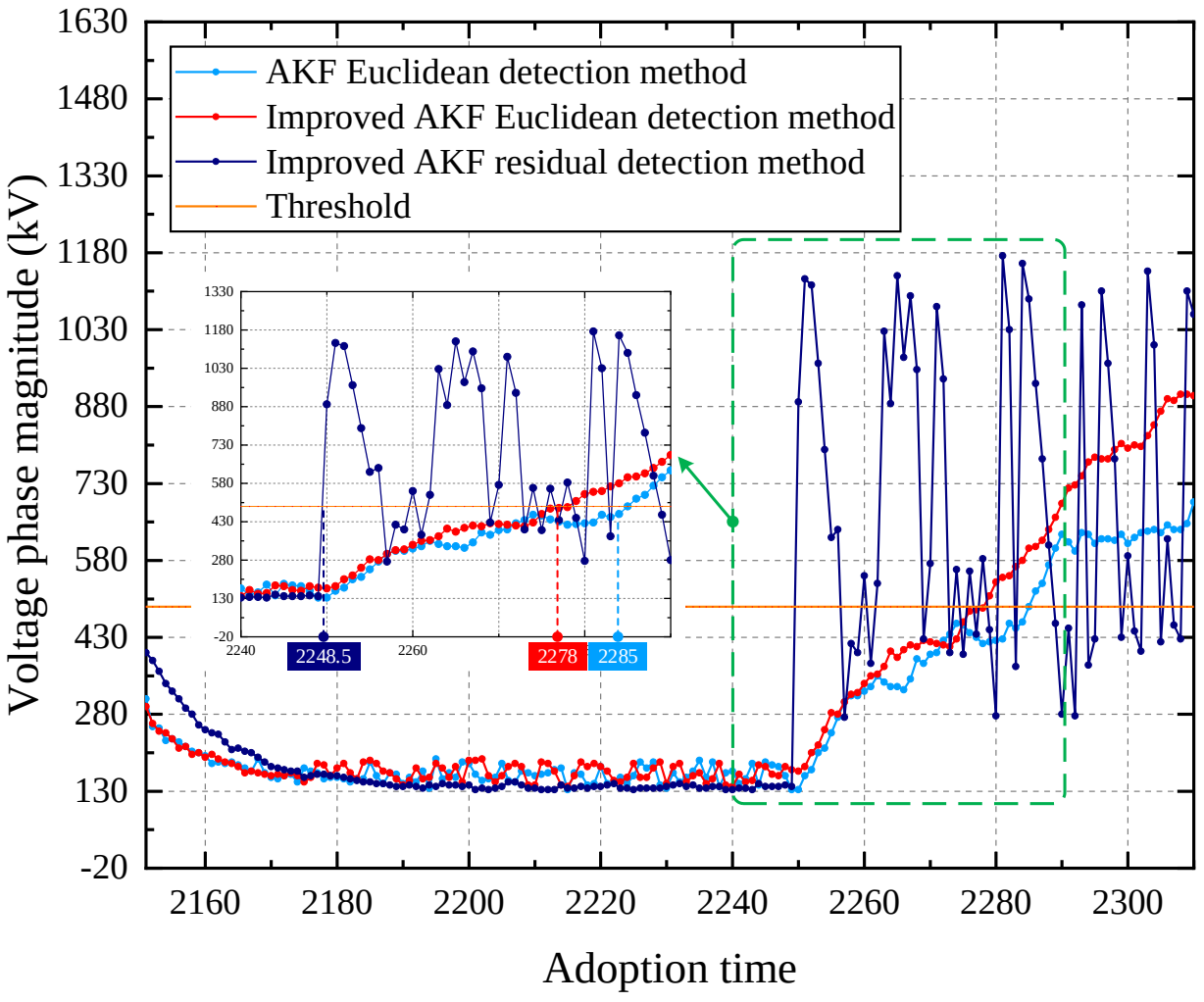

**FIGURE 9.** **Comparative analysis of FDIAs passive detection under different filtering algorithms.**

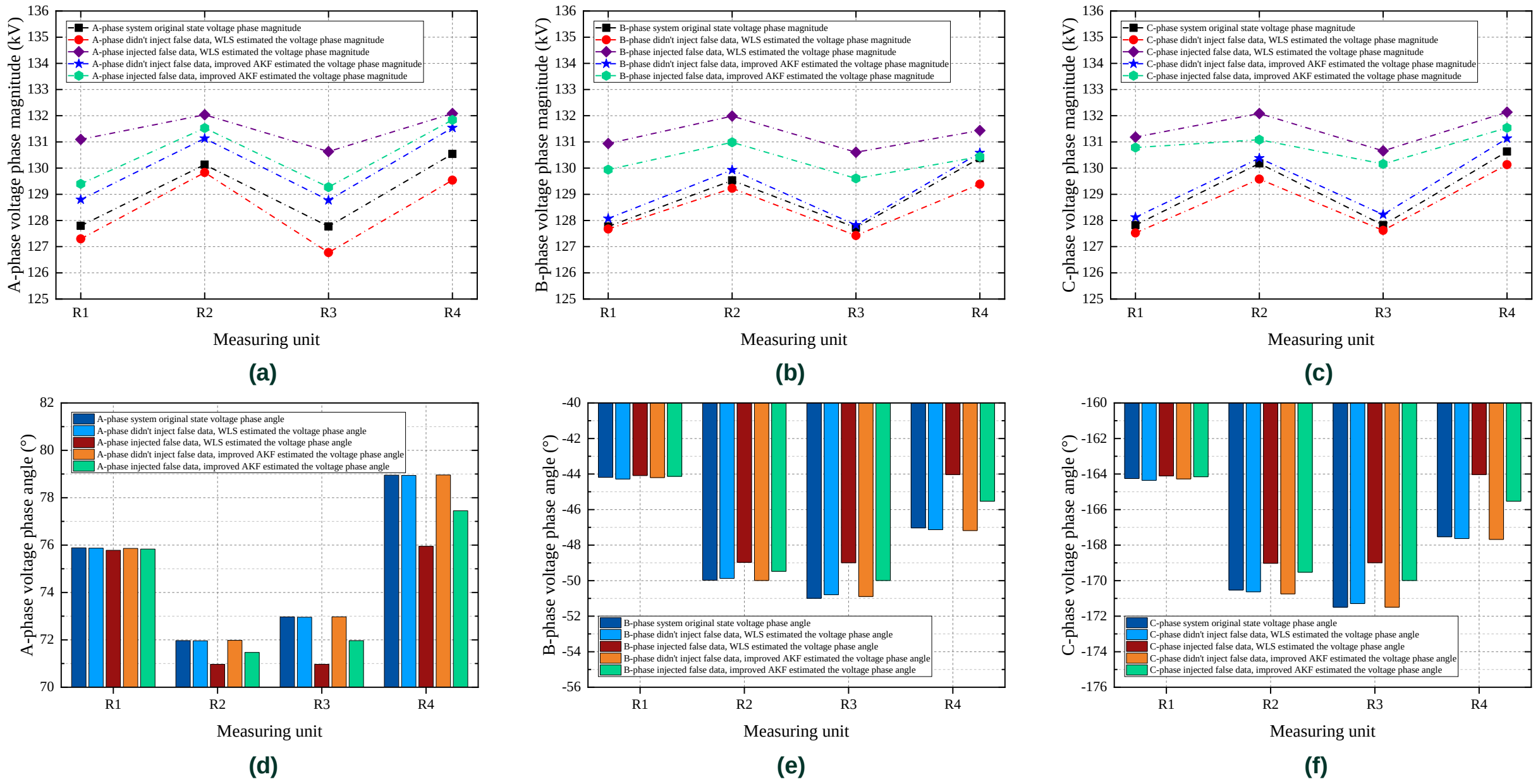

**FIGURE 10.** **R1-R4 measurement points A, B, C three-phase voltage phase magnitude, phase angle estimation results comparison.**

In Figure 10, (a)-(c) are comparison diagrams of the voltage phase magnitude of the three-phase of A, B, and C under different conditions. It can be seen from the figure that no false data is injected to improve the AKF estimated voltage phase magnitude to be closer to the original voltage. (d)-(f) are the comparison diagrams of the three-phase voltage phase angle of A, B, and C under different conditions. It can be seen from the figure that the false data is not injected to improve the AKF estimated voltage phase angle to be closer to the original voltage phase angle. The status results of R1-R4 nodes are shown in Table 1 and Table 2.

Comparing Table 1 and Table 2 before and after the attack, the voltage phase magnitude state quantity and phase angle state quantity have changed but little change. The maximum deviation of the voltage phase magnitude state quantity is 2.66kV, and the maximum deviation of the voltage phase angle state quantity is 1.22°. FDIAs has little effect on the improved AKF state estimation results. According to the statistical chi-square distribution look-up table, the bad data detection threshold is 13.34. Therefore, the improved AKF is used for state estimation, and the bad data detection system cannot distinguish false data. Then the improved AKF result is tested for residual error. The residual test formula can be calculated. When there is no false data attack vector injection attack, $r(x,\hat{x})$ =1.2723, when there is false data attack vector, $r(x,\hat{x})$ =23.6210, and the residual test threshold is set to $3\sigma$ =5.7177. It can be seen that the residual test result exceeds the threshold, and the system has FDIAs.

**TABLE 1. R1-R4 node A, B, C three-phase voltage phase magnitude status results.**

| IED number | Unattacked state/(kV) | | | Attacked state/(kV) | | | State deviation/(kV) | | |
|---|---|---|---|---|---|---|---|---|---|
| | A | B | C | A | B | C | A | B | C |
| R1 | 128.80 | 128.07 | 128.12 | 129.40 | 129.94 | 130.78 | 0.60 | 1.87 | 2.66 |
| R2 | 131.13 | 129.93 | 130.38 | 131.53 | 130.99 | 131.09 | 0.40 | 1.06 | 0.71 |
| R3 | 128.77 | 127.82 | 128.22 | 129.27 | 129.61 | 130.16 | 0.50 | 1.78 | 1.93 |
| R4 | 131.54 | 130.58 | 131.13 | 131.84 | 130.43 | 131.54 | 0.30 | 0.15 | 0.41 |

**TABLE 2. R1-R4 node A, B, C three-phase voltage phase angle status results.**

| IED number | Unattacked state/(°) | | | Attacked state/(°) | | | State deviation/(°) | | |
|---|---|---|---|---|---|---|---|---|---|
| | A | B | C | A | B | C | A | B | C |
| R1 | 75.86 | -44.20 | -164.28 | 75.83 | -44.13 | -164.16 | 0.03 | 0.07 | 0.12 |
| R2 | 71.98 | -49.99 | -170.75 | 71.47 | -49.47 | -169.53 | 0.51 | 0.52 | 1.22 |
| R3 | 72.97 | -50.89 | -171.49 | 71.97 | -49.99 | -169.99 | 1.01 | 0.90 | 1.50 |
| R4 | 78.96 | -47.18 | -167.68 | 77.45 | -45.53 | -165.53 | 1.51 | 1.65 | 2.15 |

### B. SIMULATION ANALYSIS OF FDIAS ACTIVE DETECTION BASED ON GRU-CNN

The attack dataset collected for the topology shown in Figure 8 has high power CPS characteristics and is suitable for FDIAs active detection simulation experiment verification.

This paper uses a multi-classification original dataset, considering that the purpose of FDIAs active detection is to detect whether there is FDIAs behavior from the perspective of physical measurement data mining, so the data is re-screened and annotated according to whether the categories in the data belong to FDIAs. The original data is screened and re-labeled and divided into two types of events. The first type of FDIAs event is a measurement data tampering attack, and the data is marked as "1". The second type of non-FDIAs events belong to non-attack scenarios such as the normal operation of the power system and line maintenance, and the data is marked as "0".

First, perform feature data preprocessing on each dataset that had been screened and relabeled, including missing value processing and equalization processing.

1) Since there were missing values in the original dataset, the average value method was used to complete the missing values.

2) According to statistics, there were obvious imbalances in the two types of sample data in the 15 datasets, and the number of measurement data tampering attacks was significantly lower than the normal number of samples. In this paper, the CKS oversampling algorithm designed in [39] was used to oversampling the minority samples. Figure 11 shows the distribution of normal and attacked samples in each dataset before data balancing, and Figure 12 shows the distribution of normal and attacked samples in each dataset after data balancing.

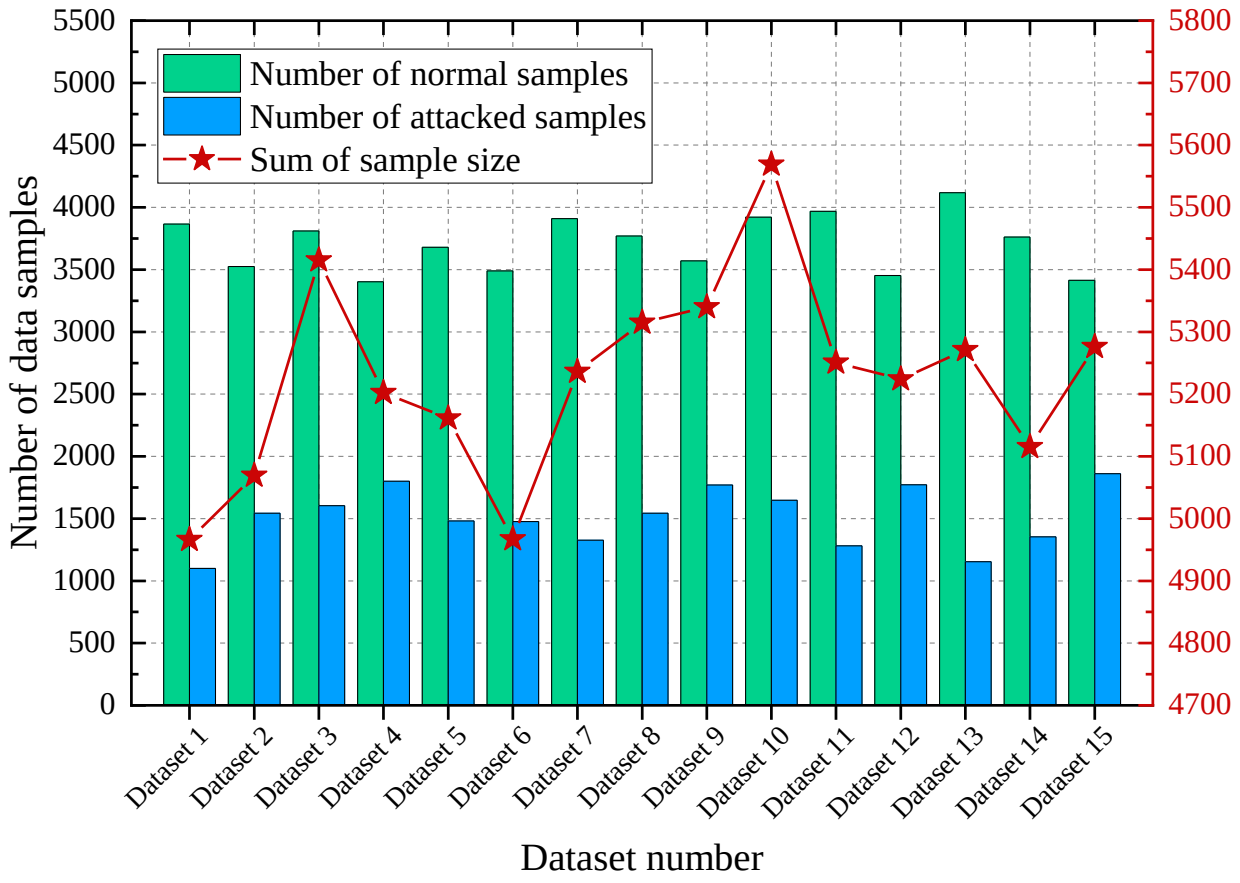

**FIGURE 11. Distribution of normal and attacked samples in each dataset before data balancing.**

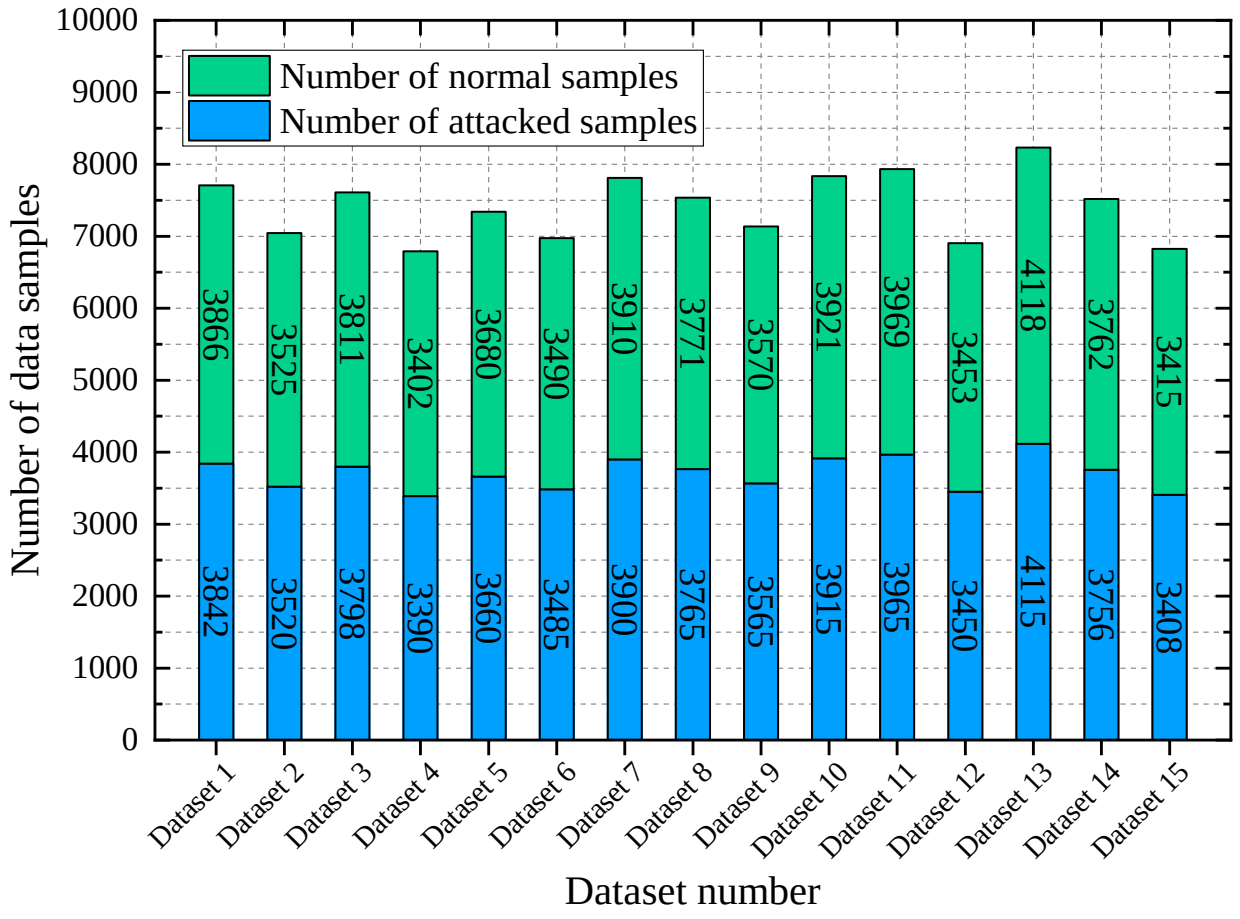

**FIGURE 12. Distribution of normal and attacked samples in each dataset after data balancing.**

Then, based on the GRU-CNN hybrid neural network structure designed in above, model training was performed, data integrity attack features were extracted, and the GRU-CNN hybrid neural network was continuously updated and adjusted according to the extraction results. Then input the test dataset into the trained GRU-CNN hybrid neural network, perform detection and classify the data through the Softmax classifier, and determine whether it was FDIAs according to the classification result. The detailed principle of the FDIAs active detection method with GRU-CNN is shown in Algorithm 1.

```
Algorithm 1: FDIAs active detection method based on GRU-CNN
  Input: Filtered and relabeled original measurement dataset
         {Dk|k∈K且K={1,2,⋯,15}}.
         IEDs collect data in real time.
  Output: FDIAs active detection results.

  Initialization:
     a).Dk, k∈K, filtered and relabeled original measurement
        dataset.
     b).DLS, large sample data set.
     c).Z, n×m matrix.
  Program:
  (Ⅰ).Obtain feature data:
     Read in the original measurement dataset Dk, k∈K.
  (Ⅱ).Basic processing of characteristic data:
     for k in K
       | Mean method to complete missing values.
       | Use CKS oversampling algorithm to generate pseudo
       | sample data and add it to Dk to achieve a balanced sample
       | number.
     end
  (Ⅲ).Feature engineering:
     Sub-sample dataset merged into large sample dataset DLS.
     Process n measured value vectors z into n×m matrix Z.
     Standardize data.
  (Ⅳ).FDIAs active detection model training:
     Large sample dataset segmentation, 80% as the training set,
     20% as the test set.
     Use 80% of the training set to train the GRU-CNN hybrid
     neural network model.
  (Ⅴ).FDIAs active detection model evaluation:
     Evaluate the GRU-CNN hybrid neural network model using
     20% of the test set.
     View metrics such as accuracy, precision, recall,
     and F1-Score.
  (Ⅵ).FDIAs active detection model application:
     Input the real-time data collected by IEDs into the FDIAs
     active detection model to determine whether it is FDIAs.
     Return FDIAs active detection results.
```

In order to verify the accuracy of the FDIAs active detection method of the GRU-CNN hybrid neural network, this paper compares GRU-CNN with other algorithms with the large sample data set DLS. These algorithms include GRU and CNN in deep learning algorithms, XGBoost in integrated learning algorithms and DT and SVM in machine learning algorithms. Figure 13 shows the active detection effect of FDIAs under different algorithm models.

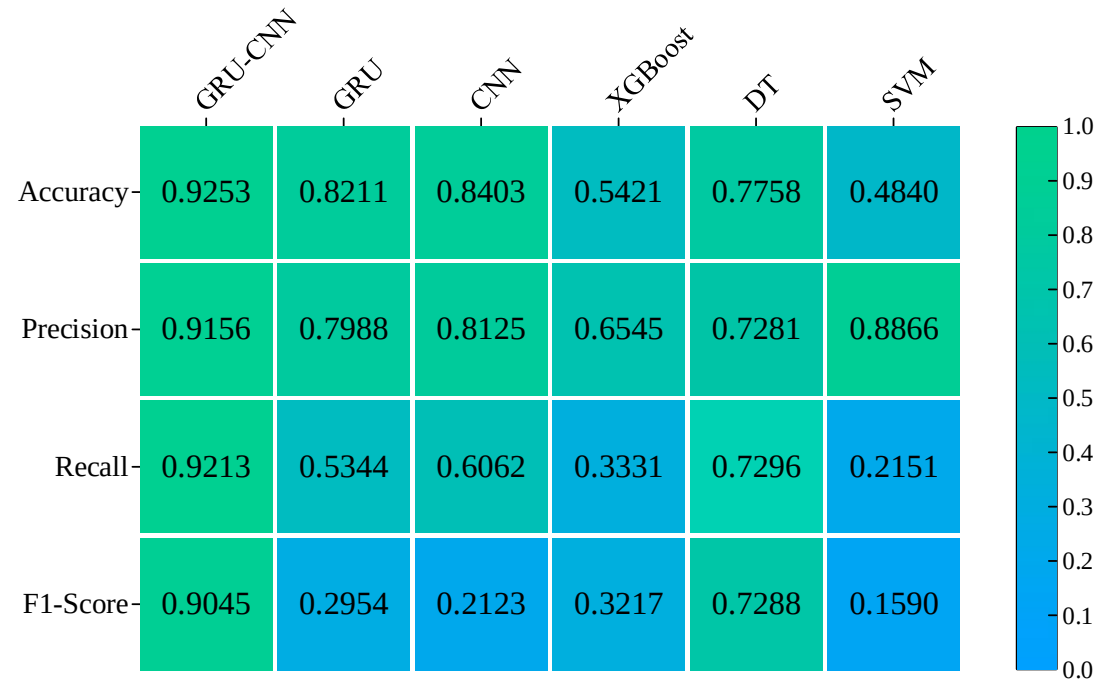

**FIGURE 13. Comparison of FDIAs active detection effect under different algorithm models.**

It can be seen from Figure 13 that the detection accuracy, precision, recall and F1-Score of the GRU-CNN model are 92.53%, 91.56%, 92.13% and 90.45% respectively. Compared with other model algorithms, the measurement indicators of detection effect have been greatly improved, the accuracy rate is increased by 44.13%, the accuracy rate is increased by 26.11%, the recall rate is increased by 70.62%, and the F1-Score is increased by 74.55%. It can be seen that the FDIAs active detection method with the GRU-CNN hybrid neural network has a better detection effect than other algorithm models.

### C. SIMULATION ANALYSIS OF ACTIVE AND PASSIVE HYBRID DETECTION

The active and passive hybrid detection method proposed in this paper belongs to the parallel mode in the typical joint mode [41]. The detection results of the knowledge-driven and data-driven methods are integrated and processed as the final output result, that is, the result of the combined operation of the two as the final output result. In order to verify the effectiveness of the active and passive hybrid detection method, with the simulation system shown in Figure 8, the voltage phase magnitude on BR3 was selected, and false data was injected at 2310 to the improved AKF passive detection, GRU-CNN active detection and active and passive hybrid detection results, as shown in Figure 14. When the output value of the detection result was 1, the system was judged as FDIAs, and when the output value was 0, the system was judged as non-FDIAs.

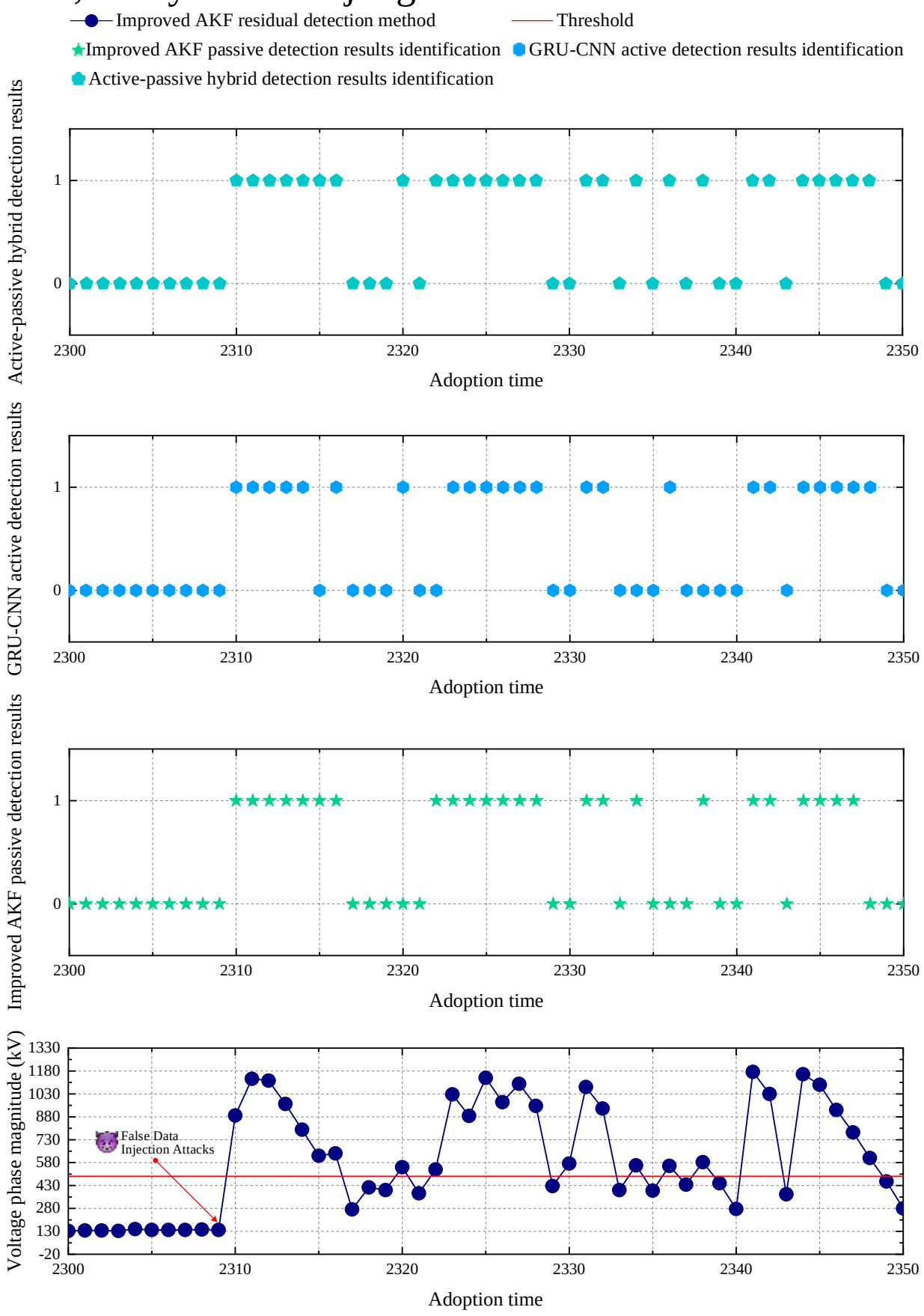

**FIGURE 14. The results of improved AKF passive detection, GRU-CNN active detection and active-passive hybrid detection.**

Analyze the effect of the proposed method through Figure 14, from time 2300 to time 2350, take the voltage phase magnitude at 50 time points for effect comparison, that is, 50 cross-section measurement values. After injecting false data at 2310, the improved AKF passive detection result showed a total of 24 cross-sections of FDIAs, the GRU-CNN active detection result showed a total of 23 cross-sections of FDIAs, and the active and passive mixed detection results showed a total of 27 cross-sections of FDIAs. Since the active and passive hybrid detection result is the combined operation result of the improved AKF passive detection and GRU-CNN active detection, the detection effect is better.

In order to verify the accuracy of the active and passive hybrid detection method (the AKF-GCNN model), this section also selects algorithms such as deep learning, ensemble learning, and machine learning to measure accuracy, precision, recall, and F1-Score. The comparison of FDIAs detection effect under different algorithm models is shown in Figure 15.

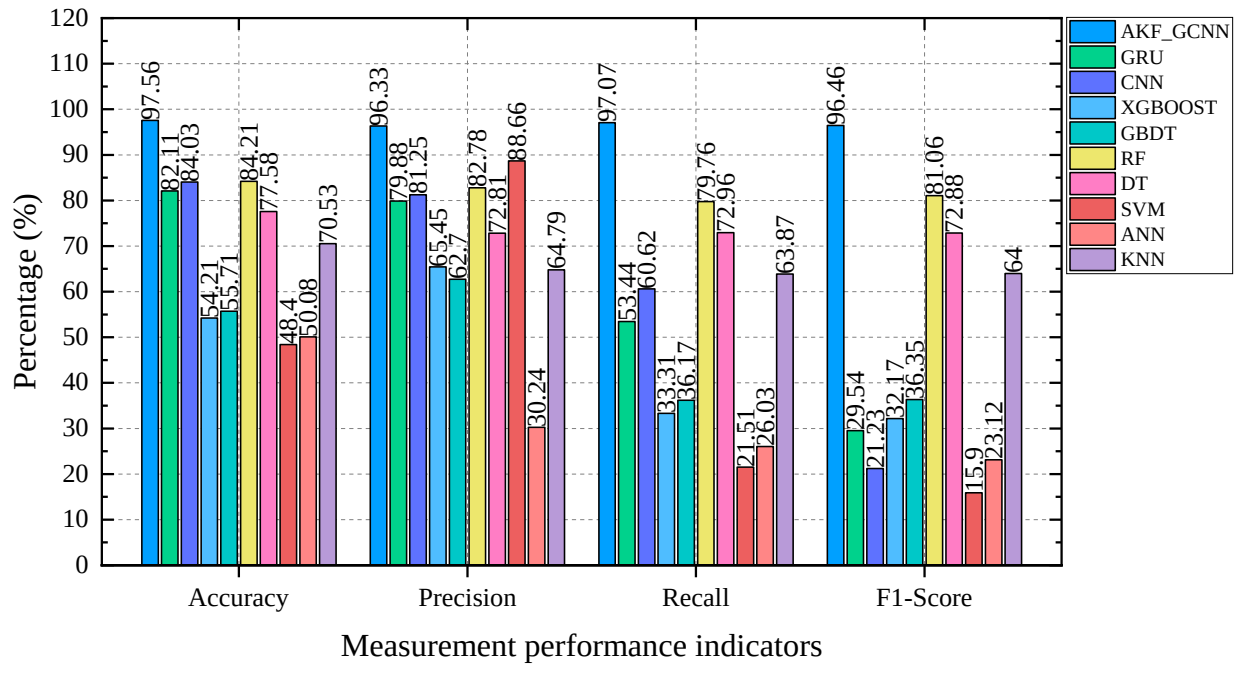

**FIGURE 15.** **Comparison of FDIAs detection effect under different algorithm models.**

It can be seen from Figure 15 that the detection accuracy, precision, recall and F1-Score of the AKF-GCNN model are 97.56%, 96.33%, 97.07%, and 96.46% respectively. Compared with other model algorithms, the measurement indicators of the detection effect have been greatly improved, the accuracy rate is up to 49.16%, the accuracy rate is up to 39.09%, the recall is up to 75.56%, and the F1-Score is up to 80.56%. It can be seen that the active and passive hybrid detection method for power CPS FDIAs with improved AKF and GRU-CNN has a better detection effect than other algorithm models.

To sum up, the FDIAs active and passive hybrid detection method proposed in this paper fully considers the deficiencies of the previous power CPS FDIAs detection methods in knowledge-driven and data-driven. The combined parallel detection mode of the two can give full play to each the advantages of each other make up for each other's shortcomings, and ultimately detect FDIAs accurately and efficiently.

## VII. CONCLUSION AND FUTURE WORK

In response to the FDIAs faced by power CPS, this paper proposes a FDIAs active and passive hybrid detection method with the improved AKF and GRU-CNN neural network, and simulates the FDIAs behavior of the power CPS. The generated operating data is used for simulation analysis. Concluded as follow:

1) The improved non-negative positive definite adaptive Kalman filter algorithm can effectively detect FDIAs by comparing the two indicators of similarity, Euclidean distance deviation and residual error, and taking into account the false alarm rate of noise.

2) The GRU-CNN hybrid neural network model is designed by combining the GRU temporal memory and the CNN spatial feature expression ability, which can effectively detect FDIAs, and the detection effect is better than the CNN neural network model.

3) The active and passive hybrid detection method with improved AKF and GRU-CNN can effectively detect FDIAs, and under the same attack dataset, the detection effect of this method is better than existing methods. At the same time, the detection method in this paper can effectively deal with the complex system environment.

Subsequent research work will further integrate the new generation of information technologies such as network attack and defense, game theory, generation and confrontation network, and consider the active defense method of the power CPS FDIAs, and then improve the adaptive immunity of the power CPS when it is attacked by the network.

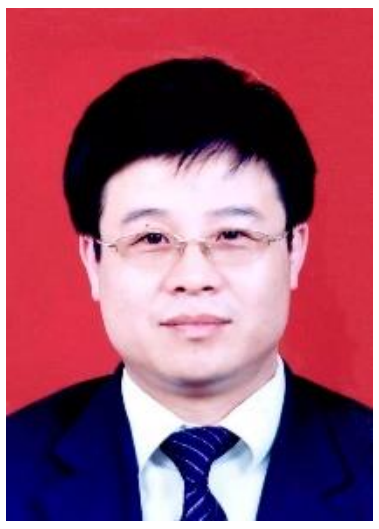

**ZHAOYANG QU** received his Ph.D. degree in Electrical Engineering from China Northeast Electric Power University in 2010, and his M.S. degree from the Dalian University of Technology in 1988. He is currently a Professor and Doctoral Tutor with the School of Computer Science, Northeast Electric Power University. He is also Vice President of the Jilin Province Image and Graphics Society, Head of the Jilin Province Engineering Technology Research Center of Power Big Data Intelligent Processing, and a Jilin Governor Baishan Scholar. His interests include network technology, smart grid, power information processing, and virtual reality. He has published more than 46 articles in SCI/EI international conference proceedings and journals. He is a member of the China Electric Engineering Society Power Information Committee. He has received the First Prize of the Jilin Province Science and Technology Progress Award. He is also a top-notch innovative talent in Jilin Province, and is designated as a young and middle-aged professional and technical talent with outstanding contributions. He has also presided over the completion of two national natural science funds.

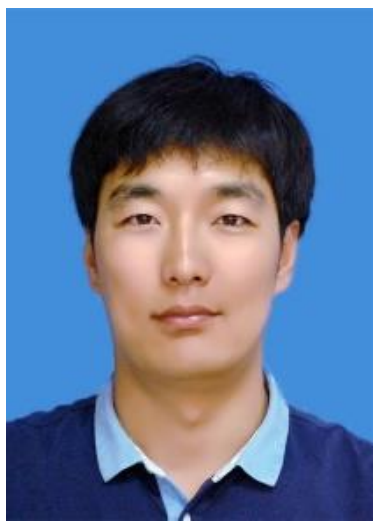

**XIAOYONG BO** received his M.S. degree in Computer Application Technology in 2014 from Northeast Electric Power University, Jilin, China, where he is currently pursuing his Ph.D. degree in Electrical Engineering. His research interests include power cyber-physical systems modeling, attack identification, and defense in smart grid.

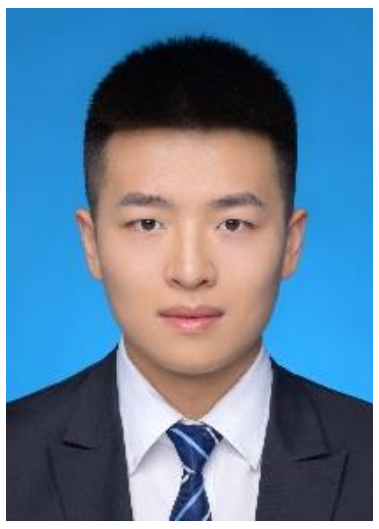

**YUNCHANG DONG** received his M.S. degree in 2019 from Northeast Electric Power University, where he is currently pursuing his Ph.D. degree in Electrical Engineering. His research interests include power cyber - physical system s and information processing in smart grid.

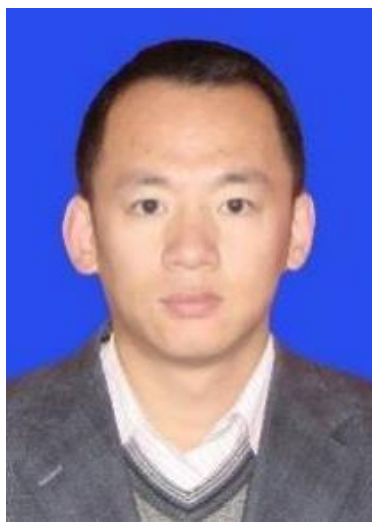

**LEI WANG** received his Ph.D. degree in Electrical Engineering from Northeast Electric Power University, Jilin, China, in 2021, where he is an Associate Professor with the School of Information Engineering. His research interests include cyber-physical systems attacks and identification in smart grid. He has received the Second Prize of the Jilin Province Science and Technology Progress Award.

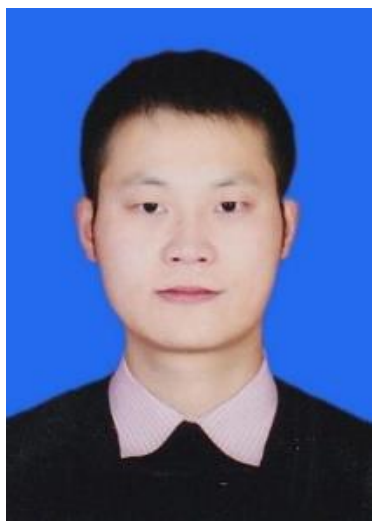

**YANG LI** received his Ph.D. degree in Electrical Engineering from North China Electric Power University, Beijing, China, in 2014. He is an Associate Professor with the School of Electrical Engineering, Northeast Electric Power University, Jilin, China. He is also a Postdoctoral Researcher with the Argonne National Laboratory, Lemont, USA, funded by the China Scholarship Council (CSC). His research interests include power system stability and control, integrated energy systems, renewable energy integration, and smart grid. He is also an Associate Editor of *IEEE Access*, *IET Renewable Power Generation*, and *Electrical Engineering*.